\documentclass[
reprint,
superscriptaddress,
nofootinbib,
 amsmath,amssymb,
 aps,
]{revtex4-2}

\usepackage[english]{babel}
\usepackage[svgnames]{xcolor}
\usepackage[linktocpage, colorlinks, citecolor=blue, linkcolor=blue, urlcolor=blue]{hyperref} 
\usepackage{enumerate}  % For enumerate environment
\usepackage{float}
\usepackage{subfigure}
\usepackage{calc}
\usepackage{dutchcal}
\usepackage{revsymb}
\usepackage{comment}
\usepackage{upgreek}
\usepackage{IEEEtrantools}
\usepackage{MyCommands}
\usepackage[normalem]{ulem}
\usepackage{csquotes}

\newcommand{\orcid}[1]{\href{https://orcid.org/#1}{\includegraphics[width=8pt]{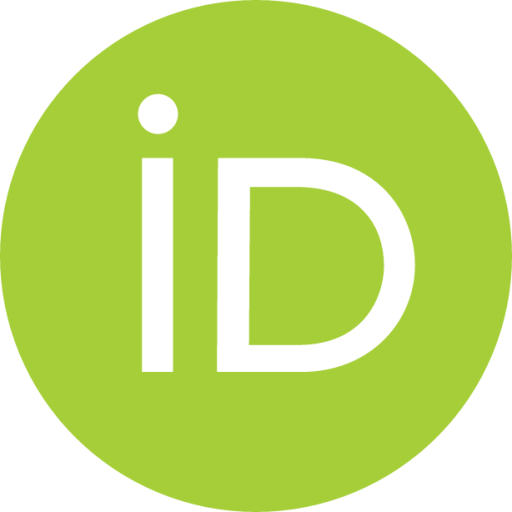}}}

\DeclareSymbolFont{cmbrightop}{OT1}{cmbr}{m}{n}
\DeclareMathSymbol{\sfPsi}{\mathalpha}{cmbrightop}{9}

\usepackage{amsfonts, amssymb, amsthm, mathtools, mathrsfs}
\usepackage{physics}

\usepackage{graphicx}% Include figure files
\usepackage{dcolumn}% Align table columns on decimal point
\usepackage{bm}% bold math

\newcommand{\gx}[1]{\textcolor{blue}{[gx: #1]}}

\begin{document}

\title{Gravitational Memory in Generalized Proca Gravity}

\author{Lavinia Heisenberg}
\affiliation{Institut f\"{u}r Theoretische Physik, Philosophenweg 16, 69120 Heidelberg, Germany}
\author{Benedetta Rosatello\orcid{0009-0001-1227-8080}}
\email{benedetta.rosatello@studenti.unipd.it}
\affiliation{Institut f\"{u}r Theoretische Physik, Philosophenweg 16, 69120 Heidelberg, Germany}
\email{benedetta.rosatello@studenti.unipd.it}
\affiliation{Dipartimento di Fisica e Astronomia “Galileo Galilei” Università di Padova, Via Marzolo 8, I-35131 Padova, Italy}
\author{Guangzi Xu\orcid{0009-0002-1276-9012}}
\email{guangzi.xu@nbi.ku.dk}
\affiliation{Institute for Theoretical Physics, ETH Z\"{u}rich, Wolfgang-Pauli-Strasse 27, 8093, Z\"{u}rich, Switzerland}
\affiliation{Center of Gravity, Niels Bohr Institute, Blegdamsvej 17, 2100 Copenhagen, Denmark}
\author{Jann Zosso\orcid{0000-0002-2671-7531}}
\email{jann.zosso@nbi.ku.dk}
\affiliation{Center of Gravity, Niels Bohr Institute, Blegdamsvej 17, 2100 Copenhagen, Denmark}
\affiliation{Albert Einstein Center, Institute for Theoretical Physics, University of Bern, Sidlerstrasse 5, 3012, Bern, Switzerland}

\begin{abstract}
We investigate the gravitational memory effect in the full Generalized Proca gravity, the most general metric theory including a gravitational Proca field with derivative self-interactions that still maintains second-order equations of motion. Building on our previous works on memory in other massless and massive metric theories, we extend a unified framework for computing displacement memory and apply it to Generalized Proca gravity.
We identify two nontrivial, physically distinct classes of background conditions of Generalized Proca theory within the assumption of asymptotic flatness: a Lorentz-invariant but massive case, and a Lorentz-violating, massless case. The former exhibits dispersive scalar and vector modes and allows a Horndeski-like treatment of memory, while the latter resembles the asymptotic dynamics of Einstein-\AE{}ther theory, including the same Lorentz-breaking effects on displacement memory.
In both cases, we derive the fully gauge-invariant and dynamical second-order action, derive the effective stress-energy tensor, and study its contribution to the memory integral. We highlight the distinction between phase and group velocity in the tensor memory formula sourced by dispersive propagating modes. Finally, we re-emphasize how observational constraints on Lorentz violation may be imposed by the structure of the memory signal.

\end{abstract}

\maketitle

%\tableofcontents

\section{\label{sec:Intro}Introduction}
What we call \emph{memory} in this work is a generic phenomenon in field theory describing a non-oscillatory component present in asymptotic radiation \cite{Zeldovich:1974gvh,Turner:1977gvh,Braginsky:1985vlg, Braginsky:1987gvh,Christodoulou:1991cr,Blanchet:1992br,Ludvigsen:1989cr,Thorne:1992sdb,PhysRevD.44.R2945,Favata:2009ii,Favata:2010zu,Bieri:2013ada,Garfinkle:2022dnm,Zosso:2025ffy,Bieri:2013hqa,Zosso:2025orw}. Memory provides a unique observational window into the notion of asymptotic symmetries \cite{Bondi:1962px,Sachs:1962wk,FrauendienerJ,Ashtekar:2014zsa,Strominger:2014pwa,Compere:2019gft,He:2014cra,Pasterski:2015zua} and the general infrared structure of field theories \cite{Weinberg_PhysRev.140.B516,Strominger:2014pwa}. 

In gravity, this effect is inevitably present in any gravitational radiation from a localized source and may soon be measured for the first time \cite{Grant:2022bla,Inchauspe:2024ibs}. Observationally, it manifests as a permanent displacement in the proper separation between freely-falling test particles following the passage of gravitational radiation. 
The dominant effect in the context of compact binary coalescences (CBCs) arises due to the nonlinearity of the field equations, or physically, from the back-reaction of the energy carried by emitted gravitational waves (GWs) \cite{Christodoulou:1991cr,Zosso:2025ffy}. While first derived within the theory of general relativity (GR) \cite{Zeldovich:1974gvh,Christodoulou:1991cr,Blanchet:1992br}, the study of gravitational memory has been recently extended to theories of gravity beyond GR \cite{lang_compact_2014,lang_compact_2015,du_gravitational_2016,koyama_testing_2020,tahura_gravitational-wave_2021,hou_gravitational_2021,tahura_brans-dicke_2021,hou_conserved_2021,hou_gravitational_2021_2,Hou:2021bxz,hou_gravitational_2022,Heisenberg:2023prj,Heisenberg:2024cjk, Heisenberg:2025tfh,Hou:2023pfz,Hou:2024exz,Zosso:2024xgy}, in which additional degrees of freedom (DOFs) most commonly described by scalar or vector fields, and symmetry-breaking mechanisms, such as Lorentz violation, can give rise to qualitatively new types of memory signals \cite{Heisenberg:2023prj, Heisenberg:2024cjk, Heisenberg:2025tfh}.
Classifying and computing gravitational memory in a broader class of gravity theories enables observational tests of GR that can distinguish between competing models and constrain existing frameworks. 

Following a novel method to understand gravitational memory from the leading-order Isaacson equations of motion, described in Refs.~\cite{Heisenberg:2023prj,Zosso:2024xgy,Zosso:2025ffy}, a unified framework for computing null and ordinary memory in scalar-tensor theories was previously developed in Ref.~\cite{Heisenberg:2024cjk}. This framework elucidated the role of dispersive scalar modes in shaping the temporal profile of the memory signal. 
Subsequently, in Ref.~\cite{Heisenberg:2025tfh}, the authors of this paper extended this analysis to Einstein-\AE{}ther (E\AE{}) gravity \cite{Jacobson:2000xp,Jacobson:2004ts,Eling:2005zq,Will:2018bme}, a manifestly local Lorentz-violating scalar-vector-tensor theory. This work derived a robust memory formula, modified by the presence of non-dispersive scalar and vector modes, which enabled a formulation of novel constraints on superluminal propagation speeds.

In this work, we compute for the first time the gravitational memory formula within Generalized Proca (GP) gravity \cite{Heisenberg:2014rta}, a general class of Ostrogradsky-stable scalar-vector–tensor models featuring derivative self-interactions and an explicit breaking of $U(1)$ gauge invariance. 
In doing so, we complete and extend a previous study of the leading-order Isaacson equations in this context \cite{Dong:2024zal}. We achieve this goal by applying and extending the fully gauge-invariant formalism and computational techniques of Refs.~\cite{Heisenberg:2023prj, Heisenberg:2024cjk, Heisenberg:2025tfh} mentioned above.

In contrast to Einstein-\AE{}ther gravity discussed in Ref.~\cite{Heisenberg:2025tfh}, which models a description of the hard breaking of local Lorentz symmetry in the gravitational sector, GP gravity does not rely on an externally imposed constraint on the vector field to avoid propagating ghost degrees of freedom. As such, the theory allows for both local Lorentz-preserving and local Lorentz-violating solutions and can therefore describe a spontaneous breaking of the latter. This work, therefore, also serves as a direct comparison between these two types of metric theories involving additional gravitational vector fields and the conceptual differences between hard and spontaneous breaking of local Lorentz symmetry.

The study of spontaneous Lorentz violation has a long tradition \cite{Nambu:1968qk,Kostelecky:1989jw,Kostelecky:1988zi,Kostelecky:2003fs,Bluhm:2004ep,Graesser:2005bg,Bluhm:2007bd,Bluhm:2007xzd}, with the most straightforward way of describing local Lorentz-breaking given by simple models of vector fields coupled to a spacetime metric, known as \emph{bumblebee models}. Generalized Proca theory can therefore be regarded as an Ostrogradsky-stable, systematic higher-order derivative extension of these models. In particular, our analysis shows that in contrast to simple bumblebee models \cite{Bluhm:2007bd,Bluhm:2007xzd}, GP gravity in a spontaneously Lorentz-breaking solution not only propagates two transverse Nambu-Goldstone (NG) vector degrees of freedom, but generally also includes an additional dynamical scalar mode.

The structure of the paper is as follows: Section~\ref{sec:basicsGP} addresses the basics of GP gravity and provides a direct comparison to the dynamics of related theories mentioned above. More specifically, after introducing the action in Sec.~\ref{ssec:action}, Sec.~\ref{ssSec:Background} starts by deriving the background equations of motion in an asymptotically flat spacetime and identifies three distinct physically relevant background solutions: $(a)$ a Lorentz-invariant case, displaying a vanishing vector background field and a nonzero hard mass for the propagating vector and scalar perturbations; $(b)$ an explicitly Lorentz-violating case, obtained by imposing a nonvanishing vector background, with massless dynamical vector and scalar degrees of freedom; $(c)$ a vanishing vector background in combination with a masslessness condition, resulting in Lorentz-preserving massless propagating modes. Since the latter case reduces to what was already discussed in Ref.~\cite{Heisenberg:2023prj}, we continue by systematically analyzing the first two cases. Following Ref.~\cite{Heisenberg:2025tfh}, we perform a fully gauge-invariant \emph{scalar-vector-tensor} (SVT) decomposition of leading-order perturbations and compute the associated second-order action [Sec.~\ref{ssSec:SecondOrderAction}]. This action allows for a systematic derivation of the equations of motion for the dynamical gauge-invariant modes for both cases $(a)$ and $(b)$, where we give particular attention to the contrast in the dispersive nature of the scalar and vector degrees of freedom between the two cases, and finally arrive in Sec.~\ref{ssSec:SecondOrderActionDynamical} at the fully gauge-invariant second-order action entirely described in terms of dynamical degrees of freedom. While this represents the first computation of this action on an asymptotically flat background, the Friedmann-Lemaître-Robertson-Walker analogue was already previously considered in Refs.~\cite{DeFelice:2016uil} and \cite{Heisenberg:2018wye}.
For completeness, we offer in Sec.~\ref{ssec:GravPol} a concise analysis of the associated gravitational polarization content of this theory, which recovers previous results in Ref.~\cite{Dong:2023xyb}.

The main content of the work is found in Sec.~\ref{sec:memory}, where the explicit computation of the gravitational memory effect in GP gravity is presented. Based on the Isaacson perturbation framework reviewed in Sec.~\ref{sSec:MemoryIsaacson} and the computation of the second-order action of purely dynamical DOFs, Sec.~\ref{ssec:emt} implements the second-variation method \cite{Maccallum:1973gf} to derive the gauge-invariant energy-momentum tensor, as previously described in Ref.~\cite{Heisenberg:2025tfh}. This result allows for the calculation of the gravitational memory effect in Section~\ref{ssec:memory}. In particular, the distinction between dispersive gauge-invariant $(a)$ and non-dispersive gauge-violating $(b)$ propagating modes motivates a careful treatment of group velocity versus phase velocity in computing the memory signal—an aspect that, to our knowledge, has not been previously addressed in this context. 
Finally, we propose observational constraints on the Lorentz-violating sector along the lines of Ref.~\cite{Heisenberg:2025tfh}. These results deepen the understanding of the impact of symmetry-breaking mechanisms on gravitational memory, and further develop the use of gravitational wave observations as a probe of beyond-GR effects.
Finally, Section~\ref{sec:Discussion} discusses the results of the present work and provides an outlook for future research directions.

We use a mostly plus $(-,+,+,+)$ metric signature, and denote spacetime indices from $0$ to $3$ by Greek letters, $\alpha,\,\beta,\,\mu,\,,\nu\,,\,...$, while spatial indices from $1$ to $3$ are denoted by Latin letters, $a,\,b,\,i,\,j,\,...$.
The symmetrization of indices is defined as $T_{(\alpha\beta)}\equiv\frac{1}{2}\left(T_{\alpha\beta}+T_{\beta\alpha}\right)$.
We choose natural units in which the speed of light $c=1$. Throughout the document, the term ``luminal'' refers to this vacuum speed of low-energy electromagnetic radiation.

%Particular attention is given to the dispersive nature of the scalar and vector degrees of freedom in the Lorentz-invariant (massive) case, in contrast to the non-dispersive behaviour exhibited in the Lorentz-violating (massless) case. 

%The former case exhibits structural similarities with the massive Horndeski theory and allows for a treatment analogous to that in \cite{Heisenberg:2024cjk}, while the latter aligns more closely with Einstein-\AE{}ther gravity as in \cite{Heisenberg:2025tfh}.

%%%%%%%%%%%%%%%%%%%%%%%%%%%%%%%%%%%%%%%%%%%%%%%%%%%%%%%%%%%%%%%%%%%%%%%%%%%%%%%%%%%%%%%%%%%%%%%%%%%%%%%%%%%%%%%%%%%%%%%%%%%%%%%%%%%%

\section{The Basics of Generalized Proca Gravity}\label{sec:basicsGP}
\subsection{Action}\label{ssec:action}
The action of generalized Proca gravity can be written as \cite{Heisenberg:2014rta,BeltranJimenez:2016rff}
\begin{eqnarray}
    S = \int \dd^4x\sqrt{-g} \sum_{i=2}^{6} \mathcal{L}_i +S_\text{m}[g,\sfPsi_\text{m}] ,
    \label{eqn:genprocaaction}
\end{eqnarray}
where
\begin{align}
    \mathcal{L}_2  =& \text{  } G_2(X,Y,F)  \text{ ,} \nonumber \\
    \mathcal{L}_3  =& \text{  } G_3(X) \nabla_\mu A^\mu  \text{ ,} \nonumber \\
    \mathcal{L}_4  =& \text{  } G_4(X)R + G_{4,X}(X)\left [ (\nabla_\mu A^\mu)^2 - \nabla_\rho A_\sigma \nabla^\sigma A^\rho \right ]  \text{ ,} \nonumber \\
    \mathcal{L}_5  =& \text{  } G_5(X)G_{\mu\nu} \nabla^\mu A^\nu 
    - \frac{1}{6}G_{5,X}(X) \left [ (\nabla_\mu A^\mu)^3 \right. \nonumber \\
     & \left. + 2\nabla_\rho A_\sigma \nabla^\gamma A^\rho \nabla^\sigma A_\gamma   
    - 3(\nabla_\mu A^\mu) \nabla_\rho A_\sigma \nabla^\sigma A^\rho  \right ] \nonumber \\
    &-\Tilde{G}_5(X) \Tilde{F}^{\alpha \mu} \Tilde{F}^{\beta}{}_\mu \nabla_\alpha A_\beta   \text{ ,} \nonumber \\
    \mathcal{L}_6  =& \text{  } G_6(X)\mathcal{L}^{\mu\nu\alpha\beta}\nabla_\mu A_\nu \nabla_\alpha A_\beta \nonumber \\
    &+ \frac{1}{2}G_{6,X}(X)\Tilde{F}^{\alpha \beta} \Tilde{F}^{\mu\nu} \nabla_\alpha A_\mu \nabla_\beta A_\nu \text{ ,}  
\end{align}
with $R$ denoting the Ricci scalar, $G_{\mu\nu}$ the Einstein tensor, $\mathcal{L}^{\mu\nu\alpha\beta} = \frac{1}{4} \epsilon^{\mu\nu\rho\sigma} \epsilon^{\alpha\beta\gamma\delta} R_{\rho\sigma\gamma\delta}$ the double dual Riemann tensor \cite{Heisenberg:2017qka} of the physical metric $g_{\mu\nu}$ that minimally and universally couples to matter fields $\sfPsi_\text{m}$ \cite{Zosso:2024xgy}. Moreover, the additional gravitational vector field $A^\mu$ defines a field strength via $F_{\mu\nu} =\nabla_{\mu} A_\nu -\nabla_{\nu} A_\mu \equiv \partial_\mu A_\nu -\partial_\nu A_\mu$ and the functions $G_i$'s are arbitrary scalar functions depending on $X\equiv -\frac{1}{2}A_\mu A^\mu$, $Y\equiv A^\mu A^\nu F_\mu{}^\alpha F_{\nu \alpha}$, $F=-\frac{1}{4}F_{\mu\nu}F^{\mu\nu}$ and $\Tilde{F}_{\mu\nu} \equiv \frac{1}{2} \epsilon_{\mu\nu\alpha\beta}F^{\alpha\beta}$ the Hodge dual of field strength $F_{\mu\nu}$, and we define their partial derivatives as $G_{i,Z} \equiv \partial G_i / \partial Z$. 
For simplicity, we assume that the scalar functions are polynomial functions. Moreover, throughout this work, we assume that $G_4\neq 0$ and $G_{2,F} \neq 0$ to ensure the dynamical nature of the metric and the Proca fields.
GP gravity is the most general metric theory extension of GR featuring an additional $U(1)$-broken vector field with derivative self-interactions that still retains an Ostrogradsky stable structure of second-order equations of motion.

%%%%%%%%%%%%%%%%%%%%%%%%%%%%%%%%%%%%%%%%%%%%%%%%%%%%%%%%%%%%%%%%%%%%%%%%%%%%%%%%%%%%%%%%%%%%%%%%%%%%%%%%%%%%%%%%%%%%%%%%%%%%%%%%%%%%

\subsection{Asymptotic perturbation theory}\label{ssec:linpert}
To study the structure of gravitational radiation emitted from a localized source within GP gravity, we consider the $\mathcal{O}(1/r)$ perturbations $\{h_{\mu\nu},\,a^\mu\}$ of the metric and the vector field, on top of an asymptotically Minkowski background metric $\eta_{\mu\nu}$ and a constant background vector field in source-centered Minkowski coordinates $\{t,r,\Omega=(\theta,\phi)\}$,
\begin{align}
   g_{\mu\nu} & = \eta_{\mu\nu}  + h_{\mu\nu}\,,\nonumber\\
    A^\mu & = \bar{A}^\mu + a^\mu\,.
\label{eqn:procadecomp}
\end{align}
Moreover, $\bar A^\mu$ is chosen to be a purely temporal constant vector field to consistently preserve the spatially homogeneous and isotropic nature of the asymptotic background
\begin{equation}\label{eq:Background}
    \bar A^\mu=(\bar{A},0,0,0)\,, \quad \nabla_\mu \bar{A}^\mu=0 \,.
\end{equation}

For simplicity, in the present work, we disregard any possible contributions of asymptotic matter fields and set  $\sfPsi_\text{m}=0$.
In the following, we will denote the vacuum metric field equation and the corresponding vector field equation of the GP action [Eq.~\eqref{eqn:procadecomp}] as
\begin{equation}
    \mathcal{G}_{\mu\nu}=0\quad\text{and}\quad \mathcal{J}_\mu=0\,,
\end{equation}
respectively. In addition, ${}_{(i)}O[a,b,\dots]$ denotes the $i$th order perturbation of the operator $O$ in the perturbation fields $\{a,b, \dots\}$  around the chosen background fields. 

In general, the perturbative degrees of freedom of the theory inherit a hard mass from the vector field $A_\mu$ whenever $\bar{G}_{2, X}\neq 0$, so that we will denote $m^2 \propto \bar{G}_{2, X}$ [Eq.~\eqref{eq:masseq}], where the value of the functions $G_i$ evaluated at the background $g_{\mu\nu}=\eta_{\mu\nu}$ and $A^\mu = \bar{A}^\mu$ is denoted as $\bar{G}_i$.

%%%%%%%%%%%%%%%%%%%%%%%%%%%%%%%%%%%%%%%%%%%%%%%%%%%%%%%%%%%%%%%%%%%%%%%%%%%%%%%%%%%%%%%%%%%
\subsubsection{Background dynamics}\label{ssSec:Background}
Given our choice of a constant background, all the terms involving derivative operators are evaluated to zero at the background level, where the equations of motion (EOMs) reduce to the simple form containing only the coefficients $\bar{G}_2$ and $\bar{G}_{2,X}$:
\begin{align}
    0=& {}_{(0)}\mathcal{J}_\mu [h_{\mu\nu},a^\mu] = \bar{A}_\mu \bar{G}_{2,X}  \,  , \label{eqn:bgproca1} \\
    0=& {}_{(0)}\mathcal{G}_{\mu\nu} [h_{\mu\nu},a^\mu] =- \frac{1}{2} \eta_{\mu\nu} \bar{G}_2 + \frac{1}{2} \bar{A}_\mu \bar{A}_\nu \bar{G}_{2,X}  \,. \label{eqn:bgproca2}
\end{align}
%Here, putting a bar on the scalar functions indicates that the functions are evaluated with the background ${\eta_{\mu\nu}, \bar{A}_\mu}$. 

%%%%%%%%%%%%%%%%%%

The background Proca equation [Eq.\eqref{eqn:bgproca1}] implies that either $\bar{A}_\mu = 0$ or $\bar{G}_{2,X}=0$. Multiplying Eq.~\eqref{eqn:bgproca1} by $\bar{A}_\nu$ and subtracting it from Eq.~\eqref{eqn:bgproca2}, the background equations imply that $\bar{G}_2=0$. Therefore, there are three distinct cases of viable background conditions in the GP theories: 
\begin{align}
   \text{(a)}&\quad \bar{G}_2=0 \,, \quad \bar{A}_\mu=0  \quad \text{and} \quad \bar{G}_{2,X}\neq 0 \,; \\
   \text{(b)}&\quad  \bar{G}_2=0 \,, \quad \bar{G}_{2,X}=0 \quad \text{and} \quad \bar{A}_\mu\neq 0\,; \\
   \text{(c)}&\quad  \bar{G}_2=0 \,, \quad \bar{G}_{2,X}=0 \quad \text{and} \quad \bar{A}_\mu = 0\,.
\end{align}
Case $(a)$ smoothly reduces to case $(c)$ as $\bar{G}_{2,X}\rightarrow 0$, while case $(b)$ is discrete, in the sense that case $(c)$ is not the $\bar A \rightarrow 0$ limit of case $(b)$. Below, we will further comment on the discreteness of this type of background solution.

The choice of the asymptotically flat background conditions above has a crucial impact on the nature of the theory. Case $(a)$ preserves local Lorentz invariance, and the degrees of freedom of the Proca field inherit a nonzero effective mass. On the other hand, case $(b)$ represents a spontaneous breakdown of local Lorentz symmetry through the presence of a nonvanishing background solution of the temporal component of the Proca field. Such a Lorentz-violation is, however, only viable in the asymptotically flat limit if the degrees of freedom are of a massless type $ \bar{G}_{2, X} \propto m^2 =0$. Note that in this case, a typical explicit potential leading to a nonzero background or ``vacuum expectation'' value in Eq.~\eqref{eq:Background} is a typical spontaneous symmetry breaking ``Mexican-hat-potential'' $G_2(X)=(X-\frac{1}{2}\bar A^2)^2$.

In the following, we will solve the asymptotic leading order equations of the theory in the non-trivial cases $(a)$ and $(b)$ and compute the corresponding gravitational memory equation of the GP theory. Since the trivial case $(c)$ is equivalent to a theory of a massless and gauge-preserving vector field non-minimally coupled to gravity discussed explicitly in Ref.~\cite{Heisenberg:2023prj}, we will not further consider it here. 
As we will show, the calculation of the memory effect in case $(a)$ is similar to the analysis of Horndeski gravity discussed in Ref.~\cite{Heisenberg:2024cjk}, while case $(b)$ is analogous to the Einstein-\AE{}ther theory treated in Ref.~\cite{Heisenberg:2025tfh}. Although in case $(a)$ it is possible to straightforwardly derive the EOMs analytically in a manifestly Lorentz preserving framework (see Appendix~\ref{App:AnalyticSolCaseA}), we will offer the same $3+1$ scalar-vector-tensor decomposition treatment for both cases in the following parts, to be able to compare the two cases directly.

%\gx{setting $m=0$ in case (a), one can check that we have the same linear equations and self-stress-energy tensor as in Ref.~\cite{Heisenberg:2023prj} \gx{replacing $A^\mu \rightarrow \nabla^\mu \Phi$, the Generalized Proca action is a special case of the SVT aciton in \cite{Heisenberg:2023prj}. the only problem is that in the other paper $\hat{G}_{4}(X) L^{\mu\nu\alpha\beta} F_{\mu\nu}F_{\alpha\beta}=4\hat{G}_{4}(X) L^{\mu\nu\alpha\beta}\nabla_\mu A_\nu \nabla_\alpha A_\beta$, which has an extra factor of $4$ comparing to the first term in $\mathcal{L}_6$ in the action we used.}

%%%%%%%%%%%%%%%%%%%%%%%%%%%%%%%%%%%%%%%%%%%%%%%%%%%%%%%%%%%%%%%%%%%%%%%%%%%%%%%%%%%%%%%%%%%%%%%%%%%%%%%%%%%%
\subsubsection{Second-order action and linear equations}\label{ssSec:SecondOrderAction}
A systematic way of deriving the leading-order dynamics of the theory is to calculate the gauge-invariant second-order action ${}_{(2)}S$ in the framework of the SVT decomposition. We refer to Appendix~\ref{app:SVTdec} for the details of this decomposition, while we recall here the 3+1 decomposition of the metric and vector field perturbations, respectively
\begin{subequations}
\begin{align}
        h_{00}=&2S,\\
        h_{0i}=&B_i^T+\partial_i B,\\
        h_{ij}=&h_{ij}^{TT}+\partial_{(i}E^T_{j)}+\left(\partial_i\partial_j-\frac{1}{3}\delta_{ij}\partial^2\right)E+\frac{1}{3}\delta_{ij}D,
\end{align}
\end{subequations}
and
\begin{equation}
    a^{\mu }=(a^{0 }, a^{T}_ i+\partial_i a).
\end{equation}
Here, $D=\delta^{ij}h_{ij}$ denotes the trace over the spatial components, the vector components satisfy the transversality condition $\partial^iB_i^T=\partial^i E_i^T=0$ and $ \partial^ia^{T}_i=0$, while the tensor component $\partial^ih_{ij}^{TT}=\delta^{ij}h^{TT}_{ij}=0$ is a transverse-traceless tensor.
Due to a well known gauge redundancy, only 10 out of the 14 degrees of freedom in this decomposition are physical. We therefore identify 10 manifestly gauge-invariant degrees of freedom, encoded within two transverse vector fields, four scalar fields
\begin{subequations}\label{eq:defgaugeinv}
\begin{align}
    \Phi&\equiv S -\dot{B}+\frac{1}{2}\ddot{E},\\
    \Theta&\equiv \frac{1}{3}(D-\partial^2E),\label{eq:Theta}\\
    \Xi_i&\equiv B_i^T-\frac{1}{2}E_i^T,\\
     \Omega &\equiv  a^{0} - \bar{A}\left(\dot{B}-\frac{1}{2}\ddot{E}\right) \,, \\
   \Upsilon &\equiv  a +\frac{1}{2}\bar{A} \dot{E} \,, \\
     \Sigma_i& \equiv  a^T_i + \frac{1}{2}\bar{A}\dot{E}^T_i\,,
\end{align}
\end{subequations}
and one transverse-traceless tensor field $h^{TT}_{ij}$, 
where an overdot denotes a derivative with respect to the time coordinate $t$.

In terms of these variables, it is then possible to write down a manifestly gauge-invariant second-order action
\begin{equation}\label{eq:SecondOrderAction}
    {}_{\myst{(2)}}S=\frac{1}{2\kappa_0}\int d^4x\sqrt{-\eta}\Big({}_{\myst{(2)}}\mathcal{L}_\text{T}+{}_{\myst{(2)}}\mathcal{L}_\text{V}+{}_{\myst{(2)}}\mathcal{L}_\text{S}\Big)\,,
\end{equation}
where the scalar, vector and tensor field contributions described by ${}_{\myst{(2)}}\mathcal{L}_\text{T},{}_{\myst{(2)}}\mathcal{L}_\text{V},{}_{\myst{(2)}}\mathcal{L}_\text{S}$ naturally decompose into distinct sectors. We present here the explicit expressions for the second-order Lagrangians that encompass both background solution cases \emph{(a)} and \emph{(b)} of interest by imposing the common background conditions,  
\begin{align}
    \bar{G}_2=0\quad \text{and}\quad \bar{A}\,\bar{G}_{2,X}=0\,.
\end{align} 
The final result reads
%\gx{the general case contains non-gauge-inv terms if only set $\bar{G}_2=0$}
\begin{subequations}\label{eq:general_action}
\begin{align}
    {}_{\myst{(2)}}\mathcal{L}_\text{T}&=-\frac{1}{4}(\bar{G}_4-\bar{A}^2\bar{G}_{4,X}){h}_{TT}^{ab}\ddot{h}^{TT}_{ab}+\frac{1}{4}\bar{G}_4 h_{TT}^{ab}\nabla^2 h^{TT}_{ab}\,,\\
{}_{\myst{(2)}}\mathcal{L}_\text{V}
    &=\bar{G}_{2,F}(\frac{1}{2}\bar{A}^2 \Xi^i\Box\Xi_i+\bar{A}\Xi^i \Box\Sigma_i+\frac{1}{2}\Sigma^i\Box \Sigma_i)\nonumber\\
    & -\bar{G}_{2,Y}(\bar{A}^4 \Xi^i \ddot \Xi_i+2\bar{A}^3 \Xi^i \ddot \Sigma_i + \bar{A}^2 \Sigma^i \ddot \Sigma_i) \nonumber \\
    &-\frac{1}{2}\bar{G}_{2,X}\Sigma^i\Sigma_i -\frac{1}{2}(\bar{G}_4+\bar{A}^2\bar{G}_{4,X})\Xi^i\nabla^2\Xi_i \nonumber \\
    &- \bar{A}\bar{G}_{4,X}\Xi^i\nabla^2\Sigma_i \,,
    \end{align}
    \begin{align}\label{eq:general_action scalar}
    {}_{\myst{(2)}}\mathcal{L}_\text{S}&=
    \bar{G}_4 (\frac{3}{2}\Theta \ddot{\Theta}-\frac{1}{2}\Theta\nabla^2\Theta +2\Phi\nabla^2\Theta) \nonumber\\
    &+\bar{G}_{4,X}[\bar{A}^2\Theta(2\nabla^2\Phi-\frac{3}{2}\ddot{\Theta})-2\bar{A}\Theta(\nabla^2\Omega+\nabla^2\dot{\Upsilon})] \nonumber\\
    &+\bar{G}_{3,X} [\frac{3}{2}\bar{A}^3\Theta\dot{\Phi}-\bar{A}^2(\frac{3}{2}\Theta\dot{\Omega}+\Upsilon\nabla^2\Phi)+\bar{A}\Upsilon\nabla^2 \Omega]\nonumber\\
    &+G_{2,XX}(\frac{1}{2}\bar{A}^4\Phi^2 -\bar{A}^3\Phi\Omega +\frac{1}{2}\bar{A}^2\Omega^2) \nonumber\\
    &+\frac{1}{2}G_{2,X}(\Omega^2+\Upsilon\nabla^2\Upsilon)\nonumber\\
    & +(\bar{G}_{2,F}+2\bar{A}^2\bar{G}_{2,Y})(-\frac{1}{2} \dot{\Upsilon}\nabla^2 \dot{\Upsilon}-\frac{1}{2}\Omega\nabla^2 \Omega\nonumber\\
    & -\dot \Upsilon \nabla^2 \Omega -2\bar{A}^2 \Phi\nabla^2 \Phi + 2\bar{A}\Omega\nabla^2\Phi+2\bar{A}\dot{\Upsilon}\nabla^2\Phi) \,.
\end{align}
\end{subequations}
Here $\Box := \eta^{\mu\nu}\partial\mu \partial\nu= -\partial_t^2+\nabla^2$ denotes the flat-space d'Alembert operator.

The next step is to vary this action with respect to the gauge-invariant scalar, vector and tensor fields to obtain the corresponding linear equations in each of the background cases $(a)$ and $(b)$. We refer to Appendix~\ref{app:perteq} for the explicit calculations, and only present here the resulting equations.

\paragraph*{Case (a).}
In this case, the leading order dynamics of GP gravity very much resemble the structure in a standard Proca theory coupled to gravity. In addition to the two transverse and massless graviton tensor modes $h^{TT}_{ij}$, the vector field contributes with the expected three additional massive dynamical degrees of freedom, which can be described in terms of the two gauge-invariant transverse vector modes $\Sigma_i$ and the gauge-invariant longitudinal scalar mode $\Upsilon$.
Indeed, the linear equation of motion for the tensor mode $h^{TT}_{ij}$ can be reduced to a massless wave equation, while the vector modes $\Sigma_i$ and the scalar mode $\Upsilon$ satisfy Klein-Gordon equations of motion, respectively
\begin{subequations}
\begin{align}
    \Box h^{TT}_{ij}&=0\,,\label{eqn:tensoreq_a}\\
    (\Box - m^2)\Sigma_i&=0 \,,\label{eqn:vectoreq_a}\\
    (\Box - m^2)\Upsilon&=0 \,.\label{eqn:scalareq_a}
\end{align}
\end{subequations}
In particular, the effective mass of the vector and scalar dynamical modes is the same, and it is defined as 
\begin{align}\label{eq:masseq}
    m^2 \equiv \bar{G}_{2,X}/\bar{G}_{2,F} \,.
\end{align} 

The scalar mode $\Omega$ satisfies a constraint equation, 
\begin{align}\label{eqn:a_constr2}
    \nabla^2 \Upsilon +\dot \Omega = 0\,,
\end{align}
which is equivalent to the Lorenz-gauge constraint $\partial_\mu a^\mu =0$ as discussed in App.~\ref{App:AnalyticSolCaseA}. While in general a combination of the two gauge-invariant variables $\Upsilon$ and $\Omega$ can be regarded as dynamical, we will always use the constraint in Eq.~\eqref{eqn:a_constr2} to choose $\Upsilon$, directly associated with the longitudinal perturbation of the vector field, as the natural dynamical quantity representing the scalar mode.\footnote{This choice is somewhat analogous to setting the temporal part of the Proca field to zero using the Lorenz-gauge constraint, as discussed in App.~\ref{App:AnalyticSolCaseA}.}

The remaining gauge-invariant modes are non-dynamical and have vanishing solutions
\begin{align}\label{eqn:a_constr1}
    \Xi_i=0\,,\quad \Theta=0\,,\quad \Phi=0\,.
\end{align}

To conclude, in the Lorentz-preserving case, the GP theory has 5 propagating degrees of freedom. The solutions to the propagation equations Eqs.~\eqref{eqn:tensoreq_a}, ~\eqref{eqn:vectoreq_a} and ~\eqref{eqn:scalareq_a} imply that the dynamical degrees of freedom can be expressed in terms of plane waves,
\begin{subequations}
\begin{align}\label{eqn:waveform}
   h^{TT}_{ij} (x) &= \frac{1}{r} f_{ij,(a)}^{TT}(t-r/V_\text{T}^{(a)},\theta,\phi)  \,,\\
   \Sigma_i (x) &= \frac{1}{r} f_{i,(a)}^{T}(t-r/V_\text{V}^{(a)},\theta,\phi)  \,,\\
   \Upsilon(x) &= \frac{1}{r} f_{(a)}(t-r/V_\text{S}^{(a)},\theta,\phi)  \,,
\end{align}
\end{subequations}
propagating along the radial direction defined by the unit vector
\begin{equation}\label{eq:n}
    n_i=(\sin\theta\cos\phi,\sin\theta\sin\phi,\cos\theta)\,,
\end{equation}
for some transverse and traceless function $f^{TT}_{ij,(a)}$, transverse function $f^{T}_{i,(a)}$ and scalar function $f_{(a)}$ respectively. These functions depend on time and radial coordinates through the combination $t-r/V_\psi^{(a)}$, for $\psi=$T, V, S, where the leading order phase velocities at $r \rightarrow \infty$ are given by
\begin{align}\label{eq:Va}
  V_\text{T}^{(a)}=1\,, \quad V_\chi^{(a)}= 1/\sqrt{1-m^2/\omega_\chi^2}
\end{align}
%$V_T=1$, and $V_\psi= 1/\sqrt{1-m^2/\omega_\psi^2}$ 
for $\chi = \text{V},\, \text{S}$. Here, $\omega_\chi$ is the energy of the corresponding wave packet, see also Appendix~\ref{app:mem_en_pulse} for a more detailed discussion.

For a non-dispersive wave packet such as the tensor mode, the group velocity, as defined in Eq.~\eqref{eq:def_velocities}, equals the phase velocity,
\begin{align}\label{eq:tensor speed a}
    \beta_{\text{T}}^{(a)} = V_\text{T}^{(a)}.
\end{align}
Thus, the corresponding asymptotic wave solution only depends on the physical retarded time $u_{\text{T}}^{(a)}\equiv t-r/\beta_\text{T}^{(a)}$, at fixed angular coordinates $(\theta, \phi)$.
\newline
On the other hand, the vector and scalar modes are dispersive, since their group velocities read, for $\chi = \text{V},\, \text{S}$,
\begin{align}\label{eq: vector and scalar  a}
    \beta_\chi^{(a)} = 1/V_\chi^{(a)} = \sqrt{1-m^2/\omega_\chi^2}\,,
\end{align}
and their dependence $t-r/V_\chi^{(a)}$ in the wave solution is not the physical velocity but rather a combination of it.

From Eq.~\eqref{eqn:waveform} we obtain a relation between the spatial derivatives and the temporal derivatives of each dynamical mode, in the asymptotic spacetime $r\rightarrow \infty$,
\begin{subequations}\label{eq:derivativestransfa}
    \begin{align}
        \partial_i h^{TT}_{kj} &= -\frac{1}{V_\text{T}^{(a)}}n_i \dot  h^{TT}_{kj} \,, \\
        \partial_i \Sigma_j &= -\frac{1}{V_\text{V}^{(a)}}n_i \dot  \Sigma_j \,, \\
        \partial_i \Upsilon &= -\frac{1}{V_\text{S}^{(a)}}n_i \dot \Upsilon \,.
    \end{align}
\end{subequations}

%%%%%%%%%%%%%%%%%%%
\paragraph*{Case (b).}
In the Lorentz-breaking case, there still exist five propagating degrees of freedom described by the following dynamical equations of motion
\begin{subequations}\label{eq:DynamicaEqsb}
\begin{align}
-\ddot h_{ij}^{TT} +V_\text{T,(b)}^2\,\partial^2h_{ij}^{TT}  &= 0\,,\label{eq:tensor_b}\\
    -\ddot \Sigma_i +V_\text{V,(b)}^2\,\partial^2\Sigma_i  &= 0\,, \label{eq:vector_b}\\
    -\ddot\Theta + V_\text{S,(b)}^2\partial^2\Theta &=0\,.
    \label{eq:scalar_b}
\end{align}
\end{subequations}
The associated phase velocities of the modes depend on the background couplings as follows
\begin{subequations}\label{eq: phase velocities b}
\begin{align}
     V_{\text{T},(b)}^2 & = \frac{\bar{G}_4}{\bar{G}_4 - \bar{A}^2\bar{G}_{4, X} } \,, \label{eq:velT}\\
   V_\text{V,(b)}^2 &=\frac{\bar{G}_{2,F}(1-\sigma \bar{A}^2)+\sigma \bar{A}^2 \bar{G}_{4,X}}{\gamma(1-\sigma \bar{A}^2)}\,, \label{eq:velV} \\
     V_\text{S,(b)}^2&= \frac{\bar{G}_{3,X}^2(-\bar{A}^2+4\bar{G}_4/\gamma)}{(1-\bar{A}^2\sigma)(3\bar{A}^2\bar{G}_{3,X}^2 + 4\bar{G}_{2,XX}\bar{G}_4(1-\sigma\bar{A}^2))}\,,\label{eq:velS}
\end{align}
\end{subequations}
where we defined the constants
\begin{subequations}
\begin{align}
    \sigma &\equiv \frac{\bar{G}_{4,X}}{\bar{G}_4}\,,\\
    \gamma&\equiv\bar{G}_{2,F}+2\bar{A}^2\bar{G}_{2,Y}\,.
\end{align}
\end{subequations}
Recall that we assumed $G_4$ to be a nonzero function for the metric to be dynamical in general, and here we further require $\gamma \neq 0$ for the Proca field to be dynamical.
In addition, as explained in Appendix~\ref{app:perteq}, these wave equations were obtained by assuming the following conditions on the background coefficients:
\begin{align}
    1-\sigma\bar{A}^2&\neq0\,,\\
    3\bar{A}^2\bar{G}^2_{3,X}+4\bar{G}_{2,XX}\bar{G}_4(1-\sigma\bar{A}^2)&\neq0\,.
\end{align}
Moreover, note that, since these equations of motion are non-dispersive, the phase velocities in Eq.~\eqref{eq: phase velocities b} correspond to the physical group velocities of the waves, which we denote as
\begin{equation}\label{eq:relPhaseGroup}
    \beta_\psi^{(b)}=V_\psi^{(b)}\,,
\end{equation}
for $\psi=$T, V, S.

There are some notable differences compared to the case $(a)$ discussed above. While the tensor sector is still governed by the gauge-invariant dynamical tensor $h_{ij}^{TT}$, it now satisfies a Lorentz-broken wave equation, which results in a departure from the propagation of the speed of light, depending on the non-minimal coupling $G_4$. The dynamics of the vector sector, too, are associated with $\Sigma_i$. Yet, compared to case $(a)$, these vector degrees of freedom lose their mass on a background solution $(b)$ and instead are governed by a Lorentz-breaking wave equation. This is very similar to what is observed in the case of Einstein-\AE{}ther gravity \cite{Heisenberg:2025tfh}. The masslessness of the vector modes in case $(b)$ can be understood by the fact that, in this case, these propagating transverse vector degrees of freedom are associated with two out of the three massless Nambu-Goldstone bosons of the spontaneously broken local Lorentz boosts \cite{Kostelecky:2003fs,Bluhm:2004ep,Bluhm:2007bd,Bluhm:2007xzd}. However, in the case of a Riemannian geometry, very generally, the possibility of an associated Higgs mechanism is not given \cite{Kostelecky:1989jw}.

The most interesting sector is the scalar sector, whose dynamics, in contrast to case $(a)$, is now entirely governed by the gauge-invariant variable $\Theta$ [Eq.~\eqref{eq:Theta}], which is solely associated with metric perturbations. Although this result is also similar to the dynamical structure of Einstein-\AE{}ther gravity, the absence of a dynamical scalar degree of freedom has a different origin in both cases. While the imposed potential in E\AE{} identically sets any massive degrees of freedom to zero, there is, in principle, a massive degree of freedom present within the Lorentz-violating branch of GP gravity, despite the condition $\bar G_{2,X}=0$. The presence of explicit mass terms is directly visible in the fourth line 
%\gx{the fourth or the fifth line?}
%\jz{It is the fourth line, since the fifth line is anyway zero in case (b)}
of the general second-order action in Eq.~\eqref{eq:general_action scalar}. However, the associated degrees of freedom turn out to be non-dynamical. This is, in fact, a known result from considerations of bumblebee models, in which the massive modes associated with the temporal component of the vector field are generally non-dynamical for constant vector backgrounds \cite{Bluhm:2007bd}. Thus, although through a different mechanism, both the local Lorentz-violating branch of GP gravity and Einstein-\AE{}ther gravity lose their dynamical massive scalar degree of freedom.
Instead, the scalar metric variable $\Theta$ acquires non-trivial dynamics, provided that $\bar G_{3,X}\neq 0$, as seen in Eq.~\eqref{eq:velS}.\footnote{Such a term is not present in typical bumblebee models considered, for instance, in \cite{Bluhm:2007bd}, which explains the absence of such a dynamical scalar mode in these theories.} These considerations highlight the fact that the background solution $(b)$ is truly distinct in nature compared to options $(a)$ and $(c)$ and, in this sense, discrete.

Similarly to case $(a)$, the asymptotic wave solutions to Eqs.~\eqref{eq:DynamicaEqsb} read
\begin{subequations}\label{eqn:waveqeb}
  \begin{align}
    h^{TT}_{ij} (x) &= \frac{1}{r} f_{ij,(b)}^{TT}(t-r/ V_\text{T}^{(b)},\theta,\phi)\,,\\
   \Sigma_i (x) &= \frac{1}{r} f^{T}_{i,(b)}(t-r/V_\text{V}^{(b)},\theta,\phi)  \,.\label{eqn:waveformV}\\
   \Theta (x) &= \frac{1}{r} f_{(b)}(t-r/V_\text{S}^{(b)},\theta,\phi)  \,.\label{eqn:waveformS}
\end{align}
\end{subequations}
This implies a similar relation between spatial and temporal derivatives, as the one found in the previous paragraph, to leading order in $r$,
\begin{subequations}\label{eq:derivativestransfb}
   \begin{align}
    \partial_i h^{TT}_{kl} &= -\frac{1}{V_\text{T}^{(b)}} n_i \dot h^{TT}_{kl}\,, \label{eq:tensor_radiation}\\
        \partial_i \Sigma_j &= -\frac{1}{V_\text{V}^{(b)}}n_i \dot \Sigma_j \,, \label{eq:vector_radiation}\\
        \partial_i \Theta &= -\frac{1}{V_\text{S}^{(b)}}n_i \dot \Theta \,.\label{eq:scalar_radiation}
\end{align}
\end{subequations}
It is interesting to consider in what parameter space the degrees of freedom of GP gravity in case $(b)$ propagate at the speed of light. Such a luminality of the tensor, vector and scalar radiations is recovered when imposing the following conditions on the background couplings:
\begin{subequations}\label{eq:luminality}
\begin{align}
    &\text{Luminality Condition 1:}\quad \bar G_{4,X}=0 \Leftrightarrow\sigma=0\,, \\ 
    &\text{Luminality Condition 2:}\quad \bar G_{2,Y}=\frac{\sigma}{2(1-\sigma\bar{A}^2)} \bar{G}_{4,X} \,, \\ 
    &\text{Luminality Condition 3:}\quad \nonumber\\
    &\bar{G}_{3,X}^2(-4\bar{A}^2+3\bar A^4 \sigma+4\bar{G}_4^2/\gamma)=4\bar G_{2,XX}\bar G_4(1-\bar A^2 \sigma)^2 \,. 
\end{align} 
\end{subequations}
In Sec.~\ref{ssec:GravPol} below we will come back to these conditions and discuss their implications on the gravitational polarizations of the theory. Moreover, as we will argue in Sec.~\ref{ssec:memory}, existing observational constraints together with the considerations of the associated gravitational memory effect might in fact favor this narrow parameter space of the theory.
On top of the five dynamical degrees of freedom discussed above, the other gauge-invariant quantities are not independent and satisfy the following constraint relations: 
\begin{subequations}
    \begin{align}
    \Xi_i &= -\sigma\bar{A} \Sigma_i\,, \label{eq:Xiconstr}\\ 
    \Phi&=\frac{\bar{G}_4}{\bar{A}^2\gamma}\Theta+\frac{1}{2\bar{A}}(\Omega+\dot\Upsilon)\,, \label{eq:phiconstr}\\ 
    \bar{G}_{2,XX}\Omega&=\bar{G}_{2,XX}\dot{\Upsilon} -\frac{2\bar{G}_{3,X}}{\bar{A}}\nabla^2\Upsilon
    +\bar{G}_{2,XX}\frac{2\bar{G}_4}{\bar{A}\gamma}\Theta\nonumber\\
     &-3\bar{G}_{3,X}\dot{\Theta}-\frac{4(\bar{G}_4-\bar{A}^2\bar{G}_{4,X})}{\bar{A}^3} \nabla^2 \Theta\,,\label{eq:omconstr}\\ 
    \bar{G}_{3,X}^2\nabla^2\Upsilon&=-\frac{2(\bar{G}_4-\bar{A}^2\bar{G}_{4,X})}{\bar{A}^2}\bar{G}_{3,X}\nabla^2\Theta\nonumber \\
    &-\left(\frac{3\bar{A}}{2}\bar{G}_{3,X}^2+\frac{2(\bar{G}_4-\bar{A}^2\bar{G}_{4,X})\bar{G}_{2,XX}}{\bar{A}}\right)\dot{\Theta}\,. \label{eq:psiconstrs}
\end{align}
\end{subequations}

%%%%%%%%%%%%%%%%%%%%%%%%%%%%%%%%%%%%%%%%%%%%%%%%%%%%%%%%%%%%%%%%%%%%%%%% 
\subsubsection{Gauge-invariant second-order Action of Dynamical Degrees of Freedom}\label{ssSec:SecondOrderActionDynamical}

The final step is to simplify the second-order gauge-invariant action using the linear relations, leaving only the dynamical degrees of freedom. In the end, the physics of the asymptotically flat vacuum in both cases depends only on five dynamical degrees of freedom.

\paragraph*{Case (a).}
As a consequence of the constraint relations among the gauge-invariant variables, Eqs.\eqref{eqn:a_constr2} and ~\eqref{eqn:a_constr1}, the second-order action reduces to 
\begin{align}\label{gaugeS2a}
{}_{\myst{(2)}} S&= 
\,\frac{1}{8\kappa_0}\int  d^4x \sqrt{-\eta}
\Bigg\{\bar C_\text{T}^{(a)}\left(\dot{h}^{TT}_{ab}\dot{h}_{TT}^{ab}-\partial_c h^{TT}_{ab}\partial^c h_{TT}^{ab}\right) \nonumber\\
&+\bar C_\text{V}^{(a)}\left(\dot{\Sigma}_a\dot{\Sigma}^a-\partial_b\Sigma_a\partial^b\Sigma^a - m^2 \Sigma_a \Sigma^a\right) \nonumber\\
&+\bar C_\text{S}^{(a)}\left(\partial_a\dot{\Upsilon}\partial^a\dot{\Upsilon}-\partial_b \partial_a \Upsilon \partial^b \partial^a \Upsilon  -m^2\partial_a\Upsilon \partial^a\Upsilon
\right)\Bigg\}\,,
\end{align}
% \textcolor{red}{
% \begin{align}
% {}_{\myst{(2)}} S=
% \,\frac{1}{8\kappa_0}\int & d^4x \sqrt{-\eta}
% \Bigg\{\bar C_\text{T}\left(\dot{h}^{TT}_{ab}\dot{h}_{TT}^{ab}-\partial_c h^{TT}_{ab}\partial^c h_{TT}^{ab}\right) \nonumber\\
% &\quad+\bar C_\text{V}\left(\dot{\Sigma}_a\dot{\Sigma}^a-\partial_b\Sigma_a\partial^b\Sigma^a - m^2 \Sigma_a \Sigma^a\right) \nonumber\\
% &\quad+\bar C_\text{S}\left(-\dot{\Omega}^2+\partial_a\Omega\partial^a\Omega +m^2\Omega^2 \right\nonumber\\
% &\left \quad +\partial_a\dot{\Upsilon}\partial^a\dot{\Upsilon}-(\nabla^2\Upsilon)^2 -m^2\partial_a\Upsilon \partial^a\Upsilon
% \right)\Bigg\}\,,
% \end{align}
% }
with 
\begin{subequations}\label{eqn:Ca}
    \begin{align}
    \bar C_\text{T}^{(a)}&=\bar{G}_4\,,\label{eq:CTa}\\
    \bar C_\text{V}^{(a)}&=2\bar G_{2,F}\,,\label{eq:CVa}\\
    \bar C_\text{S}^{(a)}&=2\bar G_{2,F}\,. \label{eq:CSa} 
\end{align}
\end{subequations}
From this action, it is easy to recover the linear equations of motion, Eqs.~\eqref{eqn:tensoreq_a}, ~\eqref{eqn:vectoreq_a}, ~\eqref{eqn:scalareq_a}. 

\paragraph*{Case (b).}
The constraint Eqs.~\eqref{eq:Xiconstr}, \eqref{eq:phiconstr}, \eqref{eq:omconstr} and \eqref{eq:psiconstrs} allow the cancellation of five of the ten variables within the second-order action [Eq.~\eqref{eq:SecondOrderAction}]. Together with the identifications of the following theory-dependent constants,
\begin{subequations}\label{eqn:Cb}
    \begin{align}
    \bar C_\text{T}^{(b)}&=\bar{G}_4\,,\label{eq:CT}\\
    \bar C_\text{V}^{(b)}&=2(1-\sigma \bar A^2)\left(\sigma \bar{A}^2 \bar{G}_{4,X} + \bar G_{2,F} (1-\sigma\bar A^2 ) \right)\,,\label{eq:CV}\\
    \bar C_\text{S}^{(b)}&=2\bar G_4\left( \frac{4\bar G_4}{\bar{A}^2 \gamma}-1\right)\,,\label{eq:CS} 
\end{align}
\end{subequations}
%\gx{double check $\bar C_S$.}\br{I had a wrong sign, now should be right. Thanks for spotting it!}
the second-order action in the Lorentz-breaking case reduces to a straightforward form, 
\begin{align}\label{gaugeS2b}
{}_{\myst{(2)}}S=\,\frac{1}{8\kappa_0}\int d^4x\sqrt{-\eta}&\Bigg\{\bar C_\text{T}^{(b)}\!\left(\frac{\dot{h}^{TT}_{ab}\dot{h}_{TT}^{ab}}{V_\text{T,(b)}^2}-\partial_c h^{TT}_{ab}\partial^c h_{TT}^{ab}\right) \nonumber\\
&\quad+\bar C_\text{V}^{(b)}\left(\frac{\dot{\Sigma}_a\dot{\Sigma}^a}{V_\text{V,(b)}^2}-\partial_b\Sigma_a\partial^b\Sigma^a\right) \nonumber\\
&\quad+\bar C_\text{S}^{(b)}\left(\frac{\dot{\Theta}^2}{V_\text{S,(b)}^2}-\partial_a\Theta\partial^a\Theta\right)\Bigg\}\,.
\end{align}
From this action, it is possible to derive directly the linear equations of motion, Eqs.~\eqref{eq:tensor_b}, ~\eqref{eq:vector_b}, ~\eqref{eq:scalar_b}.
%%%%%%%%%%%%%%%%%%%%%%%%%%%%%%%%%%%%%%%%%%%%%%%%%%%%%%%%%%%%%%%%%%%%%%%%%%%%%%%%%%%%%%%%%%%%%%%%
\subsection{Gravitational Polarizations in Generalized Gravity}\label{ssec:GravPol}
The universal and minimal coupling assumptions in metric theories of gravity imply that only the six independent physical components of the spacetime metric, that is, the six gravitational polarizations within $h^{TT}_{ij},\,\Xi_i,\,\Phi,\,\Theta$ of Eq.~\eqref{eq:defgaugeinv}, can directly interact with matter \cite{Eardley_PhysRevLett.30.884,Eardley_PhysRevD.8.3308}.
We refer to Section II.C of \cite{Heisenberg:2025tfh} for an exhaustive description of the six gauge-invariant gravitational polarizations in metric theories of gravity. In the present work, we briefly summarize this topic and present the results in the case of GP gravity.

In particular, in the asymptotics of a gravitational radiation event, the gravitational polarizations will modify the separation between two timelike geodesics. More precisely, gravitational polarizations are associated to the six independent components of the electric part of the linearized Riemann tensor that governs the geodesic deviation equation
\begin{align}
 \phantom{}_{\myst{(1)}}R_{0i0j} \equiv- \frac{1}{2}\ddot P_{ij}\,,
\end{align}
where the response matrix $P_{ij}$ can entirely be expressed in terms of the metric gauge-invariant components
\begin{equation}\label{eqn:Apolarization}
    P_{ij}=h_{ij}^{TT} + \frac{2}{V^2_\Xi} n_{(i} \Xi_{j)} + \delta_{ij} \Theta+\frac{2}{V_\Phi^2} n_i n_j \Phi\,.
\end{equation}
The polarization matrix can be expanded in terms of a convenient orthonormal basis (see e.g. \cite{Zosso:2024xgy})
\begin{align}
    P_{ij} \equiv\, & P_+ \,e^+_{ij} + P_\times\, e^\times_{ij} + P_u\, e^u_{ij} + P_v \,e^v_{ij}\nonumber \\
    &+ P_b\, e^b_{ij} + P_l\, e^l_{ij}  \,,\label{eq:PolM 2}
\end{align}
in which the six polarizations are given by
\begin{align}\label{eq:Polarization Modes}
P_+  &=  \frac{e_+^{ij}}{2}h^{TT}_{ij}\,, & P_u  &= \frac{u^i}{V_\Xi} \Xi_i\,, &P_l  &= \Theta + \frac{2}{V_\Phi^2} \Phi \,,\nonumber \\
P_\times  &= \frac{e_\times^{ij}}{2}h^{TT}_{ij}\,, & P_v  &= \frac{v^i}{V_\Xi} \Xi_i\,, &P_b &= \Theta \,.
\end{align}
We have defined here the standard transverse basis vectors $u_i$ and $v_i$ associated to a given radial direction defined in Eq.~\eqref{eq:n}
\begin{subequations}\label{eq:Def Transverse Vectors u v}
\begin{align}
u_i&=(\cos\theta \cos\phi,\,\cos\theta \sin\phi,\,-\sin\theta )\,,\\
v_i&=(-\sin\phi,\,\cos\phi,\,0)\,.
\end{align}
\end{subequations}

 At this point, we want to emphasize the key distinction between gravitational polarizations and degrees of freedom. As previously discussed, the former correspond to the physical components of the metric and are thus limited to a maximum of six. In contrast, the latter may be arbitrarily numerous, depending on the field content of the theory under consideration. The presence of additional gravitational fields can switch on some of these polarizations, in accordance with the linear dynamics of the theory. Consequently, although related, the concepts of DOFs and gravitational polarizations should not be confused. In particular, absence of additional polarizations does not directly imply the absence of any additional gravitational degrees of freedom.

To find the gravitational polarizations in the GP theory, we apply the linear equations of motion and constraints to Eqs.~\eqref{eq:Polarization Modes} and express everything in terms of the dynamical variables of the theory.

\paragraph*{Case (a).}
The gravitational polarizations in the Lorentz-invariant, massive case read
\begin{subequations}\label{eq:polarizations_a}
\begin{align}
P_+  &=  \frac{e^{ij}_+}{2}h^{TT}_{ij}\,,\\
P_\times  &= \frac{e^{ij}_\times}{2}h^{TT}_{ij}\,,\\
 %P_u  &= \sqrt{1-\frac{m^2}{\omega_\Xi^2}}\Sigma_i u^i\,, \\
 %P_v  &= \sqrt{1-\frac{m^2}{\omega_\Xi^2}}\Sigma_i v^i\,, \\
 P_u  &=0 \,, \\
 P_v  &=0 \,, \\
 \quad P_l  &= 0\,,\\
 P_b& = 0\,.
\end{align}
\end{subequations}
We observe that the scalar and vector polarizations vanish identically in this case, so that in comparison to GR no additional gravitational polarizations can be observed. %The magnitude of the vector polarization modes depends on the coefficient of this theory (the effective mass $m^2$) and the energy of the wave package $\omega_\Xi$.

\paragraph*{Case (b).}
We use the linear dynamics of the Lorentz-violating case to obtain the following polarizations,
\begin{subequations}\label{eq:polarizations}

\begin{align}
P_+  &=  \frac{e^{ij}_+}{2}h^{TT}_{ij}\,,\\
P_\times  &= \frac{e^{ij}_\times}{2}h^{TT}_{ij}\,,\\
 P_u  %&=-\sqrt{\frac{\gamma(1-\sigma \bar{A}^2)\sigma^2 \bar A^2}{\bar G_{2,F}(1-\sigma \bar{A}^2)+\sigma^2 \bar{A}^2 \bar{G}_{4}}} \Sigma_iu^i \nonumber\\
 &= -\sigma \bar A \sqrt{\frac{\gamma}{\bar G_{2,F}+\sigma^2 \bar{A}^2 \bar{G}_{4}V_{\text{T},(b)}^2}} \Sigma_iu^i \,, \\
 P_v &= -\sigma \bar A \sqrt{\frac{\gamma}{\bar G_{2,F}+\sigma^2 \bar{A}^2 \bar{G}_{4}V_{\text{T},(b)}^2}} \Sigma_i v^i\,, \\
 \quad P_l  &= \frac{1+\text{V}_{\text{S},(b)}^2(1-\bar{A}^2\sigma)-4\bar G_{4,X}/\gamma}{\text{V}_{\text{S},(b)}^2(1-\bar{A}^2\sigma)} \Theta\,,\\
 P_b& = \Theta\,.
\end{align}
\end{subequations}

The five dynamical degrees of freedom of the massless, Lorentz-violating theory excite, in general, all six gravitational polarizations. In particular, the breathing mode is always excited, the vector modes are only excited if
\begin{align}
    \sigma\neq0 \quad \text{and}\quad \gamma\neq0\,,
\end{align}
and the longitudinal one is present only if 
\begin{align}
     1+\text{V}_{\text{S},(b)}^2(1-\bar{A}^2\sigma)-4\bar G_{4,X}/\gamma\neq0\,.
\end{align}

When the luminality conditions in Eq.~\eqref{eq:luminality} are imposed, the vector and scalar polarizations reduces to
\begin{subequations}
\begin{align}
 P_u  &=0\,, \\
 P_v  &= 0\,, \\
 \quad P_l  &= 2P_b=2\Theta\,.
\end{align}
\end{subequations}
To conclude, as we previously obtained in Einstein-\AE{}ther theory ~\cite{Heisenberg:2025tfh}, the existence of vector polarizations in the Lorentz-violating sub-case of perturbed GP gravity implies that the tensor mode must propagate at a speed different from that of light, with the vector amplitude scaling proportionally to the deviation from luminal propagation via the coefficient $\sigma$. In contrast, the scalar polarizations do not exhibit the same dependence on non-luminality.

%%%%%%%%%%%%%%%%%%%%%%%%%%%%%%%%%%%%%%%%%%%%%%%%%%%%%%%%%%%%%%%%%%%%%%%%%%%%%%%%%%%%%%%%%%%%%%%%%%%%%%%%%%%%%%%%%%%%%%%%%%%%%

%\subsection{Massless linear Equations of Motion}

%%%%%%%%%%%%%%%%%%%%%%%%%%%%%%%%%%%%%%%%%%%%%%%%%%%%%%%%%%%%%%%%%%%%%%%%%%%%%%%%%%%%%%%%%%%%%%%%%%%%%%%%%%%%

%\subsection{Massless \br{gauge invariant} gauge-invariant \br{second order} second-order Action of Dynamical Degrees of Freedom}

%%%%%%%%%%%%%%%%%%%%%%%%%%%%%%%%%%%%%%%%%%%%%%%%%%%%%%%%%%%%%%%%%%%%%%%%%%%%%%%%%%%%%%%%%%%%%%%%%%%%%%%%%%%%%%%%%%%%%%%%%%%%%%%%%%%%
\section{\label{sec:Memory}Memory}\label{sec:memory}
Based on the framework developed in Refs.~\cite{Zosso:2024xgy, Heisenberg:2023prj, Heisenberg:2024cjk,Heisenberg:2025tfh}, we are now all set to analyze the nonlinear memory effect for both cases $(a)$ and $(b)$. These computations will in particular extend the previous treatment in Ref.~\cite{Heisenberg:2025tfh} - which focused on a metric theory whose asymptotic radiation has non-dispersive, constant velocities - to include a dispersive generalization.

\subsection{Leading order Isaacson equations}\label{sSec:MemoryIsaacson}
The basis of the following computation will be given by the formulation of leading order Isaacson equations of motion \cite{Isaacson_PhysRev.166.1263,Isaacson_PhysRev.166.1272,Heisenberg:2023prj,Zosso:2025ffy}. Since this approach to leading order equations of motion was previously described in great detail \cite{misner_gravitation_1973,Flanagan:2005yc,maggiore2008gravitational,Heisenberg:2023prj,Heisenberg:2024cjk,Zosso:2024xgy,Zosso:2025ffy,Heisenberg:2025tfh} we will only provide here the minimal amount of information. For the interested reader, we refer in particular to Sec. III B in Ref.~\cite{Heisenberg:2025tfh}. 

The Isaacson approach to leading-order gravitational perturbations starts with a fundamental assumption of separation of scales. Requiring the separation $f_L \ll f_H$ between the slowly varying background field of frequencies below $f_L$ and the propagating high-frequency GW signal of characteristic $f_H$, we promote the constant background metric and vector field to include a slowly varying perturbation,
\begin{align}
    \eta_{\mu\nu} &\rightarrow \eta_{\mu\nu} + \delta h^L_{\mu\nu} \,, \\
    A^\mu &\rightarrow \bar A^\mu + \delta a^\mu_L \,.
\end{align}
The subscripts $L$ and $H$ denote the low- and high-frequency perturbations.
Next, we can perform a multiscale expansion,
\begin{align}
    g_{\mu\nu} &= \eta_{\mu\nu} + \delta h^L_{\mu\nu} + h^H_{\mu\nu} \,, \\
    A^\mu &= \bar A^\mu + \delta a^\mu_L + a^\mu_H \,,
\end{align}
where the high-frequency parts have scale $|h^H|\sim |a^H| \sim \mathcal{O}(\epsilon_1)$, and the low-frequency background perturbation is of scale $|h^L|\sim |a^L| \sim \mathcal{O}(\epsilon_2)$. To validate our perturbation theory, we further impose 
\begin{align}
    \epsilon_2 \sim \epsilon_1^2 \frac{f_H^2}{f_L^2} \ll 1 \,.
\end{align}

With these assumptions in hand, one can obtain two sets of leading-order equations. The first set is the leading order equations at the high-frequency scales, 
\begin{align}
    0=&{}_{(1)} \mathcal{J}_{\mu}  [h^H,a^H] \,, \\
    0=&{}_{(1)} \mathcal{G}_{\mu\nu}[h^H,a^H] \,,\label{eq:leadingorderH}
\end{align}
which coincide in form with the linear equations obtained directly from the second-order action in the previous section. The second set describes leading order equations at the low-frequency scales, which, in terms of the metric, can be written as
\begin{align}
    {}_{(1)} \mathcal{G}_{\mu\nu} [h^L,a^L]= -\frac{1}{2}\langle {}_{(2)} \mathcal{G}_{\mu\nu} [h^H,a^H] \rangle \,,
\end{align}
where $\langle O\rangle$ defines a spacetime averaging operator that filters out the high-frequency contribution in any operator $O$ \cite{Isaacson_PhysRev.166.1263,Isaacson_PhysRev.166.1272,misner_gravitation_1973,Flanagan:2005yc,maggiore2008gravitational,Zalaletdinov:2004wd,Stein:2010pn,Favata:2011qi,Heisenberg:2023prj,Zosso:2024xgy,Zosso:2025ffy}. Note that at the low frequency scales, the operator on the right-hand-side crucially contributes to the leading order such that the equation can be interpreted as a back-reaction equation for the low frequency metric perturbation. As such, the averaged second-order operator over high-frequency fields is interpreted as the stress-energy tensor of high-frequency gravitational waves \cite{Isaacson_PhysRev.166.1263,Isaacson_PhysRev.166.1272}
\begin{align}\label{eq:EMtensor}
    {}_{(2)}t_{\mu\nu}[h^H,a^H] := -\frac{1}{2\kappa_0} \langle {}_{(2)} \mathcal{G}_{\mu\nu} [h^H,a^H] \rangle \,.
\end{align}

In general, the dynamical low-frequency contribution on the left-hand side of Eq.~\eqref{eq:EMtensor} carries the same dynamical structure as the leading-order high-frequency equation [Eq.~\eqref{eq:leadingorderH}], which has the following general form for both cases $(a)$ and $(b)$
\begin{align}
    \left[{}_{(1)} \mathcal{G}_{ij} [h,a]\right]^{TT} = \frac{1}{2}\left(-\frac{1}{V_\text{T}^2} \ddot h_{ij}^{TT} + \partial^2 h_{ij}^{TT}\right) \,.
\end{align}
The general form of the tensor back-reaction equation reads
\begin{align}\label{eq:MemoryEquation}
   \left( -\frac{1}{V_\text{T}^2} \partial_0^2 + \nabla^2 \right) \delta h_{L ij}^{TT} = 2\kappa_0 \perp_{ij,kl} {}_{(2)}t_{kl}[h^H,a^H] \,,
\end{align}
with $\perp_{ij,kl}$ denoting the standard transverse-traceless projector. As derived in Refs.~\cite{Heisenberg:2023prj,Heisenberg:2024cjk,Zosso:2025ffy,Heisenberg:2025tfh} the solution for the low-frequency perturbations of this equation provides a general formula for the associated non-linear displacement memory. This is therefore the \emph{memory equation} that we ultimately want to solve.

In order to solve this memory equation, we thus need to compute the asymptotic energy-momentum tensors [Eq.~\eqref{eq:EMtensor}] of GP gravity. It is at this stage that the effort in computing the purely dynamical and manifestly gauge-invariant second-order action in Sec.~\ref{ssSec:SecondOrderActionDynamical} pays off. This is because, while the computation of Eq.~\eqref{eq:EMtensor} in terms of gauge-invariant fields through the second-order equations of motion represents a very tedious task, the so-called second-variation approach \cite{Maccallum:1973gf} allows for a much more direct access to the information of the energy-momentum tensor, as we will explicitly show in the next section.

\subsection{Gauge-invariant Energy-Momentum tensor}\label{ssec:emt}
The asymptotic self-stress-energy tensor, although formally being an operator at second-order in perturbation theory, actually only depends on the second-order action of a theory. This statement is based on the second-variation approach, which employs a second-order action $_{\mys{(2)}} S_\text{eff}$, in which the background fields and perturbations are treated as independent fields, such that the information on the energy-momentum tensor can be accessed through variations with respect to the background fields \cite{Maccallum:1973gf,Heisenberg:2023prj,Heisenberg:2025tfh}. In particular, in the asymptotically flat limit, it was proved in Ref.~\cite{Heisenberg:2023prj} that the asymptotic energy-momentum tensor in Eq.~\eqref{eq:EMtensor} can be computed through
\begin{equation}\label{eq:lem}
    {}_{\mys{(2)}}t_{\mu\nu}[h^H,a^H]=\frac{-2}{\sqrt{-\eta}} \Biggl\langle \frac{\delta_{\mys{(2)}}S_\text{eff}^{\mys{flat}}}{\delta \eta^{\mu\nu}}\Biggr\rangle\,,
\end{equation}
where $S_\text{eff}^{\mys{flat}}$ is the Minkowskian limit of $_{\mys{(2)}} S_\text{eff}$.

To obtain the spatial-spatial part of the tensor, we use the following relation
\begin{equation}\label{SV2}
    {}_{\mys{(2)}}t_{ij}[h^H,a^H]= -2 \Biggl\langle \frac{\delta_{\mys{(2)}}S_\text{eff}^{\mys{flat}}}{\delta \eta^{ij}}\Biggr\rangle\,,
\end{equation}
in which we can directly plug in our second-order 3+1 decomposed gauge-invariant action from Eq.~\eqref{eq:SecondOrderAction}.

Since in the second-order actions  [Eqs.~\eqref{gaugeS2a} and \eqref{gaugeS2b}] each sector has the form
\begin{align}
    \mathcal{L}_\psi = \frac{\bar{C}_\psi}{8\kappa_0} \sum _\psi \left((\dot \psi)^2 -\partial_i \psi \partial^i \psi -m^2 \psi^2\right) \,,
\end{align}
for $\psi=$ T, V, S, the corresponding energy-momentum tensors after simplifying with the leading wave equation of the high-frequency fields $\psi$, should schematically read
\begin{align}
    _{\mys{(2)}}t^\psi_{ij} = \frac{1}{4\kappa_0} \sum_\psi \bar C_\psi \langle \partial_i \psi \partial_j \psi \rangle \,.
\end{align}
Note that the mass terms do not survive the averaging after imposing the high-frequency equations of motion.
The asymptotic wave solutions further imply through Eqs.~\eqref{eq:derivativestransfa} and \eqref{eq:derivativestransfb} that the energy-momentum tensor can be written as
\begin{align}
\label{eq:EmGen2}
    {}_{\mys{(2)}}t_{ij}^\psi
    =&\frac{1}{4\kappa_0}\sum_\psi \frac{\bar C_\psi}{V_\psi^2}\,\Big\langle\dot\psi_H^2\Big\rangle\,n_in_j \,.
\end{align}
 The element $t_{ij}$ in an energy-momentum tensor represents the momentum flux of momentum $i$ in the direction $j$. Since the tensor ${}_{(2)}t_{ij}$ is symmetric in $i,j$, and we only need to consider purely radial outward fluxes within the asymptotic region, it can be characterized by a single momentum flux scalar function $F_\psi$ per unit solid angle $d\Omega$ through
\begin{equation}\label{eq:EmGenFinal}
     {}_{\mys{(2)}}t^\psi_{ij}(t,r,\Omega)\equiv\frac{1}{r^2}F_\psi(u_\psi,\Omega)n_i(\Omega)n_j(\Omega)\,,
\end{equation}
where $u_\psi$ is the retarded time of the source $\psi$,
\begin{align}\label{eq:REtardedTime}
    u_\psi \equiv t- r/\beta_\psi \,.
\end{align}
Note that, in both cases $(a)$ and $(b)$, the momentum flux crucially depends on $t$ and $r$ via the group velocity $\beta_\psi$. This is due to the very general result that physical energy-momentum always propagates with the physical group velocity \cite{PhysRev.105.1129}. In Appendix~\ref{app:mem_en_pulse}, we offer a concise understanding of this result in terms of a point-particle approximation.

\begin{comment}
\gx{energy pulse:}\textcolor{red}{ 
In general, the four momentum for a plane wave with energy $\omega$ and wave vector $\vec k$ is 
\begin{align}
    p^\mu =(\omega, kn^i)= (\omega, \frac{\omega}{V}n^i)\,,
\end{align}
and its four velocity
\begin{align}
    \beta^\mu = (1, \beta^i) \,.
\end{align}
The energy-momentum tensor in the case of an energy pulse is
\begin{align}
    t^{\mu\nu}_p (x') = p^\mu_p \beta^\nu_p \delta^{(3)}(\vec{x}'-\vec x_p) \,.
\end{align}
Then we see that
\begin{align}
    t^{00}_p (x') &= \omega_p \delta^{(3)}(\vec{x}'-\vec x_p) \\
    t^{ij}_p (x') &= \omega_p \frac{\beta_p}{V_p} \delta^{(3)}(\vec{x}'-\vec x_p)\,,
\end{align}
so that
\begin{align}
    t^{ij}_p = \frac{\beta_p}{V_p} t^{00}_p \,.
\end{align}
Combining with the relation between $t_{00}$ and the actual energy flux  $\frac{dE}{dt d\Omega}$:
\begin{align}
    t_{00} = \frac{1}{r^2 \beta} \frac{dE}{dt d\Omega} \,,
\end{align}
we arrive at 
\begin{align}
    t_{ij} = \frac{1}{r^2 V} \frac{dE}{dt d\Omega} \,.
\end{align}
}\gx{if this is correct, change the definition of $F_\psi$ in Eq 107 and 109 .}
The explicit expression of the energy flux of each mode in both cases will be displayed below.
\end{comment}

\paragraph*{Case (a).}
Substituting Eq.~\eqref{gaugeS2a} in Eq.~\eqref{SV2}, we find that the spacial components of the on-shell asymptotic self-stress-energy tensor in each sector read
\begin{subequations}\label{eq:energymomentuma}
    \begin{align}
        {}_{\mys{(2)}}t^{\text{T}}_{ij,(a)}&=\frac{\bar C_\text{T}^{(a)}}{4\kappa_0} \,\Big\langle\partial_i h^{TT}_{Hab}\partial_jh_{TT}^{Hab}\Big\rangle\,,\label{eq:EmEATa} \\
        {}_{\mys{(2)}}t^{\text{V}}_{ij,(a)}&=\frac{\bar C_\text{V}^{(a)}}{4\kappa_0}  \Big\langle\partial_i \Sigma^H_{a}\partial_j\Sigma_H^{a}\Big\rangle\,, \\
        {}_{\mys{(2)}}t^{\text{S}}_{ij,(a)}&=\frac{\bar C_\text{S}^{(a)}}{4\kappa_0}  \Big\langle \partial_i  \partial_a\Upsilon^H \partial_j \partial^a\Upsilon^H\Big\rangle \,,
    \end{align}
\end{subequations}
where we recall the definition of the coefficients $\bar C_{\psi}^{(a)}$ in Eq.~\eqref{eqn:Ca}. The corresponding energy fluxes are given by
\begin{subequations}\label{eq:flux_a}
    \begin{align}
        F^{(a)}_\text{T}(u_\text{T},\Omega)&=\frac{\bar C_\text{T}^{(a)}}{4\kappa_0}\frac{r^2}{V_\text{T,(a)}^2}\langle \dot h^{TT}_{Hab}\dot h_{TT}^{Hab}\rangle\,,\\
        F^{(a)}_\text{V}(u_\text{V},\Omega)&=\frac{\bar C_\text{V}^{(a)}}{4\kappa_0}\frac{r^2}{V_\text{V,(a)}^2}\langle \dot \Sigma^{H}_a\dot \Sigma_{H}^{a}\rangle\,,\\
        F^{(a)}_\text{S}(u_\text{S},\Omega)&=\frac{\bar C_\text{S}^{(a)}}{4\kappa_0}\frac{r^2}{V_\text{S,(a)}^2} \langle \partial_a \dot \Upsilon^{H} \partial^a \dot \Upsilon^{H}\rangle\,,\label{eq:ScalarFluxa}
    \end{align}
\end{subequations}
in which velocities are defined in Eq.~\eqref{eq:Va}.

\paragraph*{Case (b).} 
In this case, we can compute the asymptotic energy-momentum tensor via the gauge-invariant, flat and exclusively dynamical second-order action in Eq.~\eqref{gaugeS2b}:
%it is now straightforward to apply the $3+1$ decomposition to Eq.~\eqref{SV2} and compute the spacial components of the on-shell asymptotic energy-momentum tensor of E\AE{} theory in each of the sectors as
\begin{subequations}\label{eq:self_stress_energy}
\begin{align}
    {}_{\mys{(2)}}t^{\text{T}}_{ij,(b)}&=\frac{\bar C_\text{T}^{(b)}}{4\kappa_0} \,\Big\langle\partial_i h^{TT}_{Hab}\partial_jh_{TT}^{Hab}\Big\rangle\,,\label{eq:EmEATb}\\
    {}_{\mys{(2)}}t^{\text{V}}_{ij,(b)}&=\frac{\bar C_\text{V}^{(b)}}{4\kappa_0}  \Big\langle\partial_i \Sigma^H_{a}\partial_j\Sigma_H^{a}\Big\rangle\,,\\
     {}_{\mys{(2)}}t^{\text{S}}_{ij,(b)}&=\frac{\bar C_\text{S}^{(b)}}{4\kappa_0}  \Big\langle\partial_i \Theta^H\partial_j\Theta^H\Big\rangle\,,
\end{align}
\end{subequations}
where the factors $\bar C_\text{T}^{(b)}$, $\bar C_\text{V}^{(b)}$ and $\bar C_\text{S}^{(b)}$ are defined in Eq.~\eqref{eqn:Cb}.
The corresponding momentum fluxes are
\begin{subequations}\label{eq:self_stress_energyFlux}
    \begin{align}
    F^{(b)}_\text{T}(u_\text{T},\Omega)&=\frac{\bar C_\text{T}^{(b)}}{4\kappa_0}\frac{r^2}{V_\text{T,(b)}^2}\langle \dot h^{TT}_{Hab}\dot h_{TT}^{Hab}\rangle\,,\\
     F^{(b)}_\text{V}(u_\text{V},\Omega)&=\frac{\bar C_\text{V}^{(b)}}{4\kappa_0}\frac{r^2}{V_\text{V,(b)}^2}\langle \dot \Sigma^{H}_a\dot \Sigma_{H}^{a}\rangle
     \,,\\
     F^{(b)}_\text{S}(u_\text{S},\Omega)&=\frac{\bar C_\text{S}^{(b)}}{4\kappa_0}\frac{r^2}{V_\text{S,(b)}^2}\langle \dot \Theta^{H}\dot \Theta^{H}\rangle\,,\label{eq:ScalarFluxb}
\end{align}
\end{subequations}
where the velocities are given in Eqs.~\eqref{eq:velT}, \eqref{eq:velV} and \eqref{eq:velS}. Note that all the asymptotic radiations are non-dispersive in this case, such that the phase velocity is equal to the group velocity $V_\psi^{(b)}=\beta_\psi^{(b)}$.

%%%%%%%%%%%%%%%%%%%%%%%%%%%%%%%%%%%%%%%%%%%%%%%%%%%%%%%%%%%%%%%%%%%%%%%%%%%%%%%%%%%%%%%%%%%%%%%%%%%%%%%%%%%%%%%%%%%%%%%%%%%%%%%%%%%%
\subsection{Memory effect}\label{ssec:memory}
With the explicit form of all the energy-momentum tensors in terms of gauge-invariant degrees of freedom computed in the previous section, we are now ready to perform the main task of this work: solve the Isaacson memory equation [Eq.~\eqref{eq:MemoryEquation}],
\begin{align}\label{eq:MemoryEq2}
    \left(-\frac{1}{\beta_\text{T}^2}\partial_0^2 +\nabla^2\right) \delta h_{ij}^{TT} = -2 \kappa_0 \perp_{ij,kl} \sum_\psi{}_{\myst{(2)}}t^{kl}_\psi \,,
\end{align}
in the limit to null infinity, taking $r \rightarrow\infty$ while keeping the retarded time constant. Since in both cases $(a)$ and $(b)$ the phase and group velocity of the tensor modes $V_\text{T}$ and $\beta_\text{T}$ are the same since the tensor modes are non-dispersive, we will, from now on, write all the tensor velocities in terms of their physical group velocity. 

We will now solve Eq.~\eqref{eq:MemoryEq2} in the same way as presented in Refs.~\cite{Heisenberg:2023prj, Heisenberg:2024cjk, Heisenberg:2025tfh}. The general solution of this sourced wave equation has the form
\begin{align}\label{eq:ansatz}
    \delta h_{ij}^{TT}(x) = -2\kappa_0 \perp_{ij,kl} \,  \int d^{4}x' \; G(x-x') \,  {}_{\myst{(2)}}t^\psi_{kl} (x')   \,,
\end{align}
for each individual source ${}_{\myst{(2)}}t^\psi_{\mu\nu}$ in Eqs.~\eqref{eq:energymomentuma} and \eqref{eq:self_stress_energy} in both cases $(a)$ and $(b)$, 
where the $x'$ refers to the coordinates of the GW source.
Here, the appropriate retarded Green's function is
\begin{align}
    G(x-x') = -\frac{\delta (t-t' - \frac{\left | \vec{x}-\vec{x}' \right |}{\beta_\text{T}} )}{4\pi  \left | \vec{x}-\vec{x}' \right | }  \, .
\end{align}

To continue, we change the coordinates of the propagating gravitational wave and the coordinates of the source, namely $x$ and $x'$, to the spherical coordinates with light-cone retarded times adapted to the appropriate propagation speeds,
\begin{align}
    (t,x,y,z) &\rightarrow (u,r,\Omega=(\theta,\phi))\, , \nonumber\\
    (t',x',y',z') &\rightarrow (u'_\psi,r',\Omega'=(\theta',\phi')) \,.\label{eq:ChangeOfVariables}
\end{align}
Specifically, $u= t-r/\beta_\text{T}$ is the retarded time of the metric radiation with associated constant group velocity $\beta_\text{T}$ and $u'_\psi= t'-r'/\beta_\psi$ the retarded time of the source, where $\beta_\psi$ is the physical propagating speed of tensor, vector, or scalar modes, depending on which source-type we are dealing with. 
Under this transformation, the volume form in the integration becomes
\begin{align}
    d^{4}x' \rightarrow r'^{2}du'_\psi dr'd\Omega' \,.
\end{align}
Recall that we identified the propagation speeds $\beta_\psi$ back in Sec.~\ref{ssSec:SecondOrderAction} in Eqs.~\eqref{eq:Va}, \eqref{eq:tensor speed a}, \eqref{eq: vector and scalar  a} and Eqs.~\eqref{eq: phase velocities b}, \eqref{eq:relPhaseGroup} for the cases $(a)$ and $(b)$, respectively. The choice of using the retarded times, which depend on the group velocities, rather than the phase velocities, in the change of variables in Eq.~\eqref{eq:ChangeOfVariables}, is due to the fact that the energy-momentum fluxes in Eq.~\eqref{eq:ansatz} explicitly depend on the physical retarded times. This represents an important subtlety in this context, which, as discussed, reflects the fundamental statement that the outgoing momentum fluxes depend on the physical group velocity of momentum transport [Eq.~\eqref{eq:REtardedTime}].

Writing the vectors $\vec{x}$ and $\vec{x}'$ as $r\vec{n}$ and $r'\vec{n}'$, where $\vec{n}$ and $\vec{n}'$ are the directions of the detector and the source with respect to the coordinate origin, we can expand the source-to-observer distance $\left | \vec{x}-\vec{x}' \right |$ in the limit $r'\ll r$,
\begin{align}
    \left | \vec{x}-\vec{x}' \right | 
    % &= \sqrt{(\vec{x}-\vec{x}')^2}  \nonumber \\
    % &=\sqrt{r^2+r'^2-2rr'\Vec{n} \cdot \Vec{n}'} \nonumber \\
    &\simeq r-r'\Vec{n}' \cdot \Vec{n} \, . 
\end{align}
With this expansion, the asymptotic Green's function becomes
\begin{comment}
\footnote{The last step is obtained by the formula 
\begin{equation}
    \delta(g(x))= {\textstyle \sum_{i}} \delta(x-x_{i})/ \left | g'(x_i) \right |\,,
\end{equation}
where $x_i$'s are roots of $g(x)$.}
\end{comment}
\begin{align}\label{eq:Greens}
    G(x-x') &=  -\frac{ \delta \left((u-u'_\psi)-r'(\frac{1}{\beta_\psi}-\frac{\Vec{n}' \cdot \Vec{n}}{\beta_\text{T}})\right) }{4\pi (r-r'\Vec{n}' \cdot \Vec{n})} \nonumber \\
    &=  \frac{\beta_\psi \,|\mathcal{V}_\psi|\,\delta (r'-(u-u'_\psi)\beta_\psi \mathcal{V}_\psi)  }{4\pi (r-r'\Vec{n}' \cdot \Vec{n})}\,,
\end{align}
where the dimensionless ratio $\mathcal{V}_\psi$ reads
\begin{equation}\label{eq:dimensionless ratio}
    \mathcal{V}_\psi \equiv \frac{1}{1-\frac{\beta_\psi}{\beta_\text{T}} \Vec{n}' \cdot \Vec{n}}\,.
\end{equation}
Note that this factor only depends on the relative angle between the source and the observer, $\mathcal{V}_\psi (\Omega',\Omega)$. 

Substituting Eq.~\eqref{eq:EmGenFinal} and Eq.~\eqref{eq:Greens} into Eq.~\eqref{eq:ansatz},  the asymptotic memory formula sourced by a single mode $\psi$ finally reads
\begin{align}\label{eq:memoryformula}
    \delta h_{ij}^{TT}(x) 
    = & -2\kappa_0 \perp_{ij,kl} 
     \int r'^2 dr'du'_\psi d\Omega' \frac{F_\psi(u'_\psi,\Omega')n'_k n'_l}{r'^2 } \nonumber\\
    & \qquad \qquad \times 
    \frac{\beta_\psi \,|\mathcal{V}_\psi|\,\delta (r'-(u-u'_\psi)\beta_\psi \mathcal{V}_\psi)  }{4\pi (r-r'\Vec{n}' \cdot \Vec{n})} \nonumber
     \\
    %=& \frac{\kappa_0}{2\pi r} \int du' d\Omega'   H[(u-u')\mathcal{V}_\psi]  F_\psi(u',\Omega')\times \nonumber\\
    %& \qquad \qquad \times \beta_\psi \mathcal{V}_\psi (\Omega')   \perp_{ij,kl}(\Omega)\, n'_k(\Omega') n'_l(\Omega') \nonumber \\
    =& \frac{\kappa_0}{2\pi r} \int du' d\Omega'   H[(u-u')\mathcal{V}_\psi] \times \nonumber\\
    & \qquad \qquad \times F_\psi(u',\Omega') \beta_\psi [  |\mathcal{V}_\psi| n'_i n'_j]^\text{TT}\,,
\end{align}
where $H[(u-u')\mathcal{V}_\psi]$ denotes the Heaviside step function.
The appearance of the step function comes from the fact that $r'\geq 0$. Hence, when eliminating the dependency of $r'$ using the delta function, we are requiring $r=(u-u'_\psi)\beta_\psi \mathcal{V}_\psi \geq 0$. We will see that the bounds of integration of the retarded time depend on the sign of the velocity ratio $\mathcal{V}_\psi$. 

For a simpler notation, we omit the labels \emph{(a)} and \emph{(b)} in the following paragraphs, unless needed to clarify the discussion.

\paragraph*{Case (a).}
In this case, the scalar and vector modes are dispersive massive degrees of freedom, such that we always have
\begin{align}
    \beta_\chi \equiv 1/V_\chi =  \sqrt{1-m^2/\omega_\chi^2} \leq 1=\beta_\text{T} \,,
\end{align}
%$\beta_\psi \equiv 1/V_\psi =  \sqrt{1-m^2/\omega_\psi^2} \leq 1=\beta_\text{T}$ 
for $\chi =$V, S. Hence, the velocity ratio $\mathcal{V}_\psi$, for $\psi=$T, S, V, is always positive. Consequently the step function reads $H[(u-u')\mathcal{V}_\psi]=H(u-u')$ and the memory sourced by each mode of the high-frequency radiation fields with associated energy $\omega_\psi$ is 
\begin{align}\label{eq:memory_a}
    \delta h_{ij}^\text{TT} (x) = \frac{\kappa_0}{2\pi r} \int_{-\infty}^{+\infty} du' & \int_{S^2}d\Omega'  \,   H(u-u')\nonumber\\
    &\times F_\psi(u', \Omega') \beta_{\psi} \left[  \mathcal{V}_\psi n'_i n'_j\right]^\text{TT} \, \nonumber\\
    = \frac{\kappa_0}{2\pi r} \int_{-\infty}^{u} du' &\int_{S^2}d\Omega'  \, F_\psi(u', \Omega') \nonumber \\
   &\times\beta_{\psi} \left[  \mathcal{V}_\psi n'_i n'_j\right]^\text{TT}\,.
\end{align}
Notice that when $\beta_\psi^{(a)} = \beta_\text{T}^{(a)}=1$, the divergence in $\mathcal V_\psi=\mathcal V_\text{T}$ along the directions $\vec n'\cdot \vec n = 1$  is cured by the TT projection. Thus, the displacement memory remains finite for all values of the parameter space (see Ref.~\cite{Heisenberg:2025tfh} for more details). In particular the memory sourced by the tensor modes is the same as in standard GR.

The memory solution in Eq.~\eqref{eq:memory_a} is, however, only the contribution of the individual plane wave modes at given frequencies. For the scalar and vector modes with varying group velocity depending on the frequency $\omega_\chi$ of each mode, one needs to sum over all possible frequencies in order to obtain the full result. This can effectively be obtained through an integration over all possible emission speeds. Summing over all SVT sectors, the total tensor memory in GP gravity in the background case $(a)$ therefore reads 
\begin{align}
    \delta h_{ij}^\text{TT} (x) = \frac{\kappa_0}{2\pi r} \int_{-\infty}^{u} du' \int_{S^2}d\Omega' F_\text{T}(u',\Omega') \left[ \mathcal{V}_\text{T} n'_i n'_j\right]^\text{TT} \nonumber\\
    +\frac{\kappa_0}{2\pi r} \sum_{\chi = \{\text{S,V}\}} \int_0^1 d\beta_\chi 
    \int_{-\infty}^{u} du' \int_{S^2}d\Omega' F_\chi(u', \Omega') \times \nonumber\\ \times \beta_{\chi}\left[\mathcal{V}_\chi  n'_i n'_j\right]^\text{TT}  \,,
\end{align}
where in case $(a)$ the energy fluxes %$F_{\text{T}}^{(a)},\,F_\chi^{(a)}$ 
are defined in Eq.~\eqref{eq:flux_a}, 
%the phase velocities within these fluxes are given in Eq.~\eqref{eq:Va}
and the group velocities within the dimensionless ratio $\mathcal{V}_\psi$ [Eq.~\eqref{eq:dimensionless ratio}] are given in Eqs.~\eqref{eq:tensor speed a} and ~\eqref{eq: vector and scalar  a}.

\paragraph*{Case (b).}
In the Lorentz-violating case, the velocity ratio $\mathcal{V}_\psi$ is no longer guaranteed to be positive. Therefore, we will discuss the memory for different group velocities separately. 

First, recall that in this case all modes are non-dispersive, so that the group velocities equal the phase velocities of each mode, $\beta_\psi = V_\psi$. Let us then discuss two distinct possibilities. If $\beta_\text{T}\geq\beta_V$ and $\beta_\text{T}\geq\beta_S$ (possibility $(i)$), then 
\begin{subequations}
\begin{align}\label{eqn:memoryconstraint_b}
    \bar G_{2,Y} &\geq \frac{\sigma}{2}(\bar G_{4,X} -\bar G_{2,F}) \,, \\
    \bar{G}_{2,XX} &\geq \frac{\bar{G}_{3,X}^2(\bar{G}_4-\gamma\bar{A}^2)}{\gamma\bar{G}_4(1-\sigma\bar{A}^2)} \,,
\end{align} 
\end{subequations}
respectively, whereas if $\beta_\text{T}<\beta_V$, $\beta_\text{T}<\beta_S$ (possibility $(ii)$), the reverse inequalities hold. 

\emph{(i) }When $\beta_\text{T} \geq \beta_\chi$, the velocity ratio $\mathcal V_\psi$ is positive in all directions $\Omega'$. Therefore, the memory converges and is similar to case $(a)$, such that the memory solution sourced by $\psi$ in background solution $(b)$ reads
\begin{align}\label{eq:memory_b1}
    \delta h_{ij}^\text{TT} (x) = \frac{\kappa_0}{2\pi r} \int_{-\infty}^{u} du' &\int_{S^2}d\Omega'  \, F_\psi(u'_\psi, \Omega') \nonumber\\
    &\times   \beta_{\psi} \left[  \mathcal{V}_\psi n'_i n'_j\right]^\text{TT}\,.
\end{align}

\begin{comment}
    $V_T < V_V$ when 
\begin{align}
    \bar G_{2,Y} < \frac{\sigma}{2}(\bar G_{4,X} -\bar G_{2,F}) \,.
\end{align}

$V_T < V_S$ when
\begin{align}
    {\bar{G}_{3,X}^2(\bar{G}_4-\gamma\bar{A}^2)} > \gamma\bar{G}_{2,XX}\bar{G}_4(1-\sigma\bar{A}^2)\,.
\end{align}
\end{comment}

\emph{(ii) }When $ \beta_\text{T} < \beta_\psi$, the velocity ratio $\mathcal V_\psi (\Omega')$ diverges in the direction in which $\vec n' \cdot \vec n = \beta_\text{T}/\beta_\psi$ and the memory magnitude becomes formally unbounded from above. Given that the observer's location is fixed, denote this critical direction by $\Omega'=\Omega_0 = (\theta_0,\phi_0)$.
When $\vec n' \cdot \vec n \neq \beta_\text{T}/\beta_\psi$, the sign of $\mathcal V_\psi$ divides the memory formula into two separate pieces through the behaviour of the step function $H[(u-u')\mathcal V_\psi]$, which we denote as case ($+$) and case ($-$). In case ($+$), in which $\mathcal{V}_\psi(\Omega')>0 \Leftrightarrow \vec{n}\cdot\vec{n}' < \beta_\text{T}/\beta_\psi$, the memory formula still reads
\begin{align}\label{eq:memory_b2+}
    \delta h_{ij+}^\text{TT} (r,u,\Omega,\Omega') = \frac{\kappa_0}{2\pi r} \int_{-\infty}^u du'&  F_\psi(u', \Omega') \nonumber \\
    &\times \beta_\psi
   \left[ \mathcal{V}_\psi n'_i n'_j\right]^\text{TT}\,.
\end{align}
On the other hand, in case ($-$), that is, in the regime where $\mathcal{V}_\psi(\Omega')<0 \Leftrightarrow \vec{n}\cdot\vec{n}' > \beta_\text{T}/\beta_\psi$ one obtains
\begin{align}\label{eq:memory_b2-}
    \delta h_{ij-}^\text{TT}(r,u,\Omega,\Omega') = \frac{\kappa_0}{2\pi r} \int^{+\infty}_u du'&  F_\psi(u', \Omega') \nonumber \\
    &\times \beta_\psi
   \left[ -\mathcal{V}_\psi n'_i n'_j\right]^\text{TT}\,.
\end{align}
The extra dependence on $\Omega'$ of the displacement memory indicates here that the above solutions represent the memory contribution which arises from the source that is emitted along a certain direction.
In Eq.~\eqref{eq:memory_b2-}, we integrate the retarded time from $u$ to $+\infty$ because the velocity ratio $\mathcal{V}_\psi$ is negative in this case. Consequently, the integrand is nonvanishing only when $u' > u$, as required by the step function $H[(u - u')\mathcal{V}_\psi]$ in the general formula, Eq.~\eqref{eq:memoryformula}.
Physically, this arises because the projection of the velocity of the source modes $\psi$ in the direction of tensor-mode propagation is superluminal, $\beta_\psi \vec{n} \cdot \vec{n}' > \beta_\text{T}$. Along this specific spatial direction $\Omega$, the source $\psi$ therefore approaches the neighbourhood of future null infinity $\mathscr{I}^+$ earlier than the tensor mode emitted at the same time. As a result, the tensor mode that arrives at the detector at null infinity with retarded time $u_0$ can be influenced by sources emitted at any later time $u > u_0$, as one can see in Fig.~\ref{fig:penrosediagram}.

Eqs.~\eqref{eq:memory_b1}, \eqref{eq:memory_b2+} and \eqref{eq:memory_b2-} represent the general solutions of gravitational memory in the background solution case $(b)$. We recall that, in this case, the energy fluxes are defined in Eq.~\eqref{eq:self_stress_energyFlux}, while the group velocities are just phase velocities and can be found in Eq.~\eqref{eq: phase velocities b}. 
While Eq.~\eqref{eq:memory_b1} does not contain any particularly new feature of displacement memory, the case ($+$) and ($-$) in Eq.~\eqref{eq:memory_b2+} and \eqref{eq:memory_b2-} are worth analyzing further.

In the neighborhood of the critical direction $\Omega_0$, the memory is significantly enhanced. In fact, the memory becomes too large as $\Omega' \rightarrow \Omega_0$, and our assumptions of perturbation theory break down. Nevertheless, the increase in gravitational memory is physical and can be geometrically explained as the directions in which source-momentum in $\psi$ is emitted along the past null cone of a given memory event. Unlike other directions of emission, in which a source wave packet would only intersect the past light cone of the memory event once, a source particle moving along the critical direction provides a continuous contribution to the memory buildup at the same spacetime point, causing an a priori unbounded memory amplitude buildup. This phenomenon has an exact analogue within the Lorentz-breaking Einstein-\AE{}ther theory, discussed at length in Section IV.B.3 of Ref.~\cite{Heisenberg:2025tfh}. We refer the interested reader to that paper for more details and intuition on the physical explanation and expectation of the memory offset amplification.

While this phenomenon represents an interesting signal to be searched for in GW data, one can follow the same lines of reasoning as in Ref.~\cite{Heisenberg:2025tfh} and already consider first constraints on GP gravity. Indeed, provided that compact binary systems within GP gravity will in fact excite the additional GP modes within the radiation emission, the quantitative amplitude of the memory enhancement described above is already bounded by the unsuccessful search for gravitational memory within compact binary coalescences \cite{Tiwari:2021gfl,Cheung:2024zow}. It is interesting to combine this overall consideration with other existing observational bounds on GP gravity within the background solution branch $(b)$. First of all, the multimessenger observation from a binary neutron star merger constrains the propagating speed of the tensor GWs to be approximately the speed of light \cite{LIGOScientific:2017adf} up to a correction of the order of $10^{-15}$. On the other hand, the absence of a phenomenon of Cherenkov radiation \cite{Moore:2001bv} from high-energy cosmic rays effectively rules out the possibility of having subluminal propagating modes $\psi$ within the background solution $(b)$. Indeed, the same arguments used in the context of Einstein-\AE{}ther gravity \cite{Elliott_2005,Gupta:2021vdj,Sarbach:2019yso} should apply to case $(b)$ of GP gravity.
Using these two constraints, the parameter space of case $(b)$ of GP gravity is already restricted to \cite{Heisenberg:2025tfh}
\begin{align}
    \beta^{(b)}_T = 1 + \mathcal{O}(10^{-15})\,, \quad \beta^{(b)}_\psi \geq 1-\mathcal{O}(10^{-15}) \,.
\end{align}
In this context, a potentially very large memory enhancement could effectively only be avoided in the case of a restriction to an entirely luminal parameter space $\beta^{(b)}_\psi=1$ given by the conditions in Eqs.~\eqref{eq:luminality}.

\section{Discussion and Conclusion}\label{sec:Discussion}

In this work, we compute for the first time the gravitational displacement memory effect within the theory of GP gravity. We achieve this goal through the construction of the second-order perturbed action in a fully gauge-invariant form, expressed purely in terms of the physical degrees of freedom. In combination with an Isaacson perturbative framework, this action allows a systematic extraction of the leading-order dynamics of the propagating modes, which, in particular, includes the low-frequency self-sourced memory equation. 

We perform the analysis for the two main cases $(a)$ and $(b)$ of nontrivial background solutions in the asymptotically flat limit and offer a concise comparison of the two cases throughout the work. Our analysis thereby confirms the known polarization structure of the theory and yields the effective stress-energy tensor associated with the radiative modes. The main results of the work are the complete description of the leading-order gravitational wave memory formulas in both background solution branches.

The results within the first branch $(a)$ of a Lorentz preserving background are well in line with previous studies on similar massless and massive scalar-vector-tensor theories \cite{Heisenberg:2023prj, Heisenberg:2024cjk}. The additional Proca field provides three extra dynamical degrees of freedom that can appear in general asymptotic radiation. In the massive case, their propagation equations are dispersive, with a subluminal propagation speed depending on the mass and the given energy scale. The knowledge of the full memory formula consequently depends on an explicit integration over the frequency mode content of emitted vector and scalar radiation. The study of this effect on realistic gravitational memory data represents a particularly interesting future task.

On the other hand, the results in the local Lorentz-breaking branch $(b)$ recover in their essence the conclusions drawn within the manifestly Lorentz-violating theory of Einstein-\AE{}ther gravity \cite{Heisenberg:2025tfh}. The masslessness of the vector degrees of freedom can be explained through their association with the Nambu-Goldstone bosons of broken Lorentz symmetry \cite{Bluhm:2007bd}, while there also exists an additional massless scalar degree of freedom within the metric perturbations. It would be interesting to confirm whether this extra propagating degree of freedom could be associated with a massless NG boson of a broken diffeomorphism, which very generally accompanies a local violation of Lorentz symmetry \cite{Bluhm:2007xzd}.

Most importantly, the memory effect equally features intriguing amplifications along specific spacetime directions. This behaviour is tied to the causal structure defined by the relative speed between the energy-carrying source and the tensor gravitational waves. Specifically, when the source propagates faster than the tensor modes, the memory amplitude becomes unbounded at a critical observation angle where the source’s wavefront intersects the observer’s past light cone. In close analogy to the discussion in Ref.~\cite{Heisenberg:2025tfh}, this mechanism may imply strong novel bounds on the Lorentz-breaking sector of GP gravity. In a similar way to case $(a)$, a more detailed and quantitative analysis of this distinctive effect on gravitational memory represents a promising avenue for future research towards the exploration of the constraining power of the memory effect.

Although the results on gravitational memory between Einstein-\AE{}ther gravity and the Lorentz-violating solutions of GP gravity exhibit a large overlap, the difference between the theories lies in the structure of the free parameters of the theory as well as in their mechanism for avoiding the propagation of destabilizing ghost degrees of freedom. Most importantly, GP gravity allows for the possibility of describing multiple types of vacuum expectation values within the same theoretical framework. This would, in principle, allow for the consideration of randomized initial conditions in a cosmological setting, possibly exhibiting interaction effects between different regions of solutions, a direction that might be interesting to explore in the future.

Other possibilities for follow-up research are given by the investigation of an even broader space of asymptotic background solutions of GP gravity, in particular, the possibility of nonzero spatial background vector values. Along a similar line, the possibility of a Higgs mechanism accompanying the spontaneous breaking of Lorentz symmetry, which is allowed in the case of Riemann-Cartan geometries \cite{Bluhm:2007bd,Bluhm:2007xzd}, could potentially feature interesting new phenomenology within the gravitational memory effect as well.

\begin{acknowledgments}
J.Z. is supported by funding from the Swiss National Science Foundation
Grant No. 222346. The Center of Gravity is a Center of Excellence funded by the Danish National Research Foundation under grant No. 184.

\end{acknowledgments}

%_____________________________________________________________________

%%%%%%%%%%%%%%%%%%%%%%%%%%%%%%%%%%%%%%%%%%%%%%%%%%%%%%%%%%%%%%%%%%%%%%%%%%%%%%%%%%%%%%%%%%%%%%%%%%%%%%%%%%%%%%%%%%%%%%%%%%%%%%%%%%%%

\appendix

\section{Gauge-invariant SVT decompositions}\label{app:SVTdec}
In Section \ref{ssec:linpert}, we resort to the so-called gauge-invariant scalar-vector-tensor decomposition \cite{Lifshitz:1945du,Flanagan:2005yc,Zosso:2024xgy} in order to decouple the sectors within the gauge-invariant second-order action Eq.~\eqref{eq:SecondOrderAction} and derive the decoupled linearized equations of motion in the presence of spatial isotropy. We present here the details of the identification of the gauge-invariant variables.

As anticipated in Section~\ref{ssSec:SecondOrderAction}, in this framework, the components of the metric perturbation can be uniquely decomposed as $SO(3)$-invariant parts,
\begin{subequations}
\begin{align}
        h_{00}=&2S,\\
        h_{0i}=&B_i^T+\partial_i B,\\
        h_{ij}=&h_{ij}^{TT}+\partial_{(i}E^T_{j)}+\left(\partial_i\partial_j-\frac{1}{3}\delta_{ij}\partial^2\right)E+\frac{1}{3}\delta_{ij}D,\label{eq:Decombmetric}
\end{align}
\end{subequations}
where $D=\delta^{ij}h_{ij}$ is the trace over the spatial components, the components $\partial^iB_i^T=\partial^i E_i^T=0$ are transverse vector and $\partial^ih_{ij}^{TT}=\delta^{ij}h^{TT}_{ij}=0$ is a transverse-traceless tensor. On the other hand, the vector perturbation is decomposed as
\begin{equation}
    a^{\mu }=(a^{0 }, a^{T}_ i+\partial_i a),
\end{equation}
where $\partial^ia^{T}_i=0$ is a transverse vector field.

These perturbed fields are subject to unphysical gauge redundancies \cite{Nakamura:2020pre, carroll2019spacetime, Zosso:2024xgy} parametrized by an infinitesimal vector field $\xi^\mu$, 
\begin{align}
    h_{\mu\nu} &\to h_{\mu\nu}+\mathcal{L}_{\xi}\eta_{\mu\nu}= h_{\mu\nu}+\partial_\mu\xi_\nu+\partial_\nu\xi_\mu\,,\label{eq:gaugetrM}\\
    a^\mu &\to a^\mu+\mathcal{L}_{\xi}\bar{A}^\mu=a^{\mu }-\bar{A}\dot\xi^\mu\,,
    \label{eq:gaugetrA}
\end{align}
and the physical observables ought to be invariant under these transformations. The latter indicates that only $10$ out of the initial $14$ perturbation metric and vector field components can be physical degrees of freedom. To identify the gauge-invariant ones, we decompose $\xi^{\mu}$ into its temporal, transverse and longitudinal parts,
\begin{align}
\xi^{\mu}=(\xi^0, \xi^{T}_{i}+\partial_i\xi),
\end{align}
where $\partial^i\xi^{T}_{i}=0$, such that the gauge transformations in Eqs.~\eqref{eq:gaugetrM} and \eqref{eq:gaugetrA} become
\begin{subequations}
\begin{align}
     h_{ij}^{TT} &\to h_{ij}^{TT},\\
     E_i^T &\to E_i^T +2\xi_i^T,\\
    B_i^T&\to B_i^T+\dot{\xi_i},\\
    S&\to S-\dot{\xi^0},\\
    B&\to B-\xi^0+\dot{\xi},\\
    D&\to D+2\partial^2\xi,\\
    E&\to E+2\xi,\\
    a^{T}_i &\to a^{T}_i - \bar{A} \dot \xi^T_i\,,\\
    a^{0 } &\to a^{0 } - \bar{A} \dot\xi^0 \,, \\
    a &\to a -\bar{A} \dot \xi \,.
\end{align}
\end{subequations}
While the TT part of the metric $h_{ij}^{TT}$ is already gauge-invariant, the remaining $8$ gauge-invariant physical degrees of freedom can be chosen as 
\begin{subequations}\label{eq:ggaugepot}
\begin{align}
    \Phi&\equiv S -\dot{B}+\frac{1}{2}\ddot{E},\label{ggaugepot1}\\
    \Theta&\equiv \frac{1}{3}(D-\partial^2E),\\
    \Xi_i&\equiv B_i^T-\frac{1}{2}E_i^T,\\
     \Omega &\equiv  a^{0} - \bar{A}\left(\dot{B}-\frac{1}{2}\ddot{E}\right) \,, \\
   \Upsilon &\equiv  a +\frac{1}{2}\bar{A} \dot{E} \,, \\
     \Sigma_i& \equiv  a^T_i + \frac{1}{2}\bar{A}\dot{E^T_i}\,.
    \label{ggaugepot2}
\end{align}
\end{subequations}

\section{Manifestly Lorentz-invariant solution of linear EOMs in case (a)}\label{App:AnalyticSolCaseA}

In this Appendix, we present an alternative, more direct and manifestly Lorentz invariant way of computing the linear equations of motion in case $(a)$.
Observe that all the components in the action are quadratic or higher than quadratic in $A_\mu$, except for the term $\bar{G}_4 R$. Thus, all the terms in the linear metric equation are of order $\mathcal{O}(A^2)$, except $\bar{G}_4 G_{\mu\nu}$. Perturbing the equations to the first order and setting $\bar{A}_\mu=0$, all the terms of order $\mathcal{O}(A^2)$ vanish, and the linear metric equation will be
\begin{align}
    \Box h_{\mu\nu} & = 0 \,.
\end{align}
Similarly, for the vector equation, we can drop the terms of order $\mathcal{O}(A^3)$ in the action. Then we are left with only three terms: $\bar{G}_{2,X}X$, $\bar{G}_{4,X}XR$, and $\bar{G}_{4,X}[(\nabla_\mu A^\mu)^2-\nabla_\rho A_\sigma \nabla^\sigma A^\rho]$. The contribution from $\bar G_3 \nabla_\mu A^\mu$ is not considered, since it is equivalent to $A^\mu \nabla_\mu \bar G_3$ in the action through integration by parts, and the derivative of the constant $\bar G_3$ vanishes. Using this effective action to calculate the linear vector field equation, we obtain $\partial^\nu F_{\nu\mu} - m^2 a_\mu=0$, where the mass $m$ is defined in Eq.~\eqref{eq:masseq}. Contracting this equation with $\partial^\mu$, we obtain a constraint equation
\begin{align}
    \partial^\mu a_\mu =  0\,.
\end{align}
The linear vector equations can be further simplified using this constraint equation and become 
\begin{align}
    (\Box-m^2)a_\mu =0 \,.
\end{align}
Without loss of generality, we can use the constraint to set the temporal component $a^0$ to $0$. 

In summary, we obtain a wave equation for the metric field perturbation and a Klein-Gordon equation for the full vector field perturbation in a manifestly Lorentz-invariant form. The non-linear displacement memory effect is thus expected to resemble the result obtained within Horndeski theory \cite{Heisenberg:2024cjk}.
Indeed, the form of the energy-momentum tensor is analogous: applying the same strategy to the right-hand side of Eq.~\eqref{eq:lem} dropping the $\mathcal{O}(A^3)$ terms in the effective action within the asymptotic limit, and varying with respect to the background metric field, we obtain the following expression for the stress-energy tensor
\begin{align}
    {}_{\mys{(2)}}t_{ij}&=\frac{\bar G_4}{4\kappa_0} \,\Big\langle\partial_i h^{TT}_{ab}\partial_j h_{TT}^{ab}\Big\rangle
    +\frac{\bar G_{2,F}}{2\kappa_0}  \Big\langle\partial_i a_k\partial_j a^k\Big\rangle\,. 
\end{align}

%%%%%%%%%%%%%%%%%%%%%%%%%%%%%%%%%%%%%%%%%%%%%%%%%%%%%%

\section{Perturbed action and equations of motion}\label{app:perteq}
In this appendix, we describe the details of the computations of Sec.~\ref{sec:basicsGP}.
Recall that, in this section, we expand the GP action  Eq.~\eqref{eqn:genprocaaction} to the second-order and perform a 3+1 SVT decomposition. Then, we impose the background conditions to reduce the two cases \emph{(a)} and \emph{(b)} of interest, that is
\begin{align}
    \bar{G}_2=0\quad \text{and}\quad \bar{A}\,\bar{G}_{2,X}=0\,.
\end{align} 
We obtain a gauge-invariant second-order action, 
\begin{equation}
    {}_{\myst{(2)}}S=\frac{1}{2\kappa_0}\int d^4x\sqrt{-\eta}\Big({}_{\myst{(2)}}\mathcal{L}_\text{T}+{}_{\myst{(2)}}\mathcal{L}_\text{V}+{}_{\myst{(2)}}\mathcal{L}_\text{S}\Big)\,,
\end{equation}
where $\sqrt{-\eta}=1$ for a Minkowski background and ${}_{\myst{(2)}}\mathcal{L}_\text{T}$, ${}_{\myst{(2)}}\mathcal{L}_\text{V}$,${}_{\myst{(2)}}\mathcal{L}_\text{S}$ are given in Eqs.~\eqref{eq:general_action}.
%\gx{the general case contains non-gauge-inv terms if only set $\bar{G}_2=0$}
\subsection{Case (a)}
By further imposing the background condition $\bar{A}=0$, the second-order action in Eqs.~\eqref{eq:general_action} reduces to
\begin{subequations}\label{eq:a_sec_action}
\begin{align}
    _{(2)}\mathcal{L}_T &= \frac{1}{4}\bar{G}_4 h^{TT}_{ij} \Box h_{TT}^{ij} \,,\\
    _{(2)}\mathcal{L}_V &= \frac{1}{2} \Sigma_{i} (\bar{G}_{2,F}\Box-\bar{G}_{2,X}) \Sigma^{i} - \frac{1}{2}\bar{G}_4 \Xi_i\nabla^2\Xi^i \,,\\
    _{(2)}\mathcal{L}_S 
    &=\bar{G}_4 (\frac{3}{2}\Theta \ddot{\Theta}-\frac{1}{2}\Theta\nabla^2\Theta +2\Phi\nabla^2\Theta) \nonumber\\
    &+\frac{1}{2}\bar{G}_{2,X}(\Omega^2+\Upsilon\nabla^2\Upsilon) \nonumber\\
    &+ \bar{G}_{2,F}(\frac{1}{2} \ddot{\Upsilon}\nabla^2 \Upsilon-\frac{1}{2}\Omega\nabla^2 \Omega -\dot \Upsilon \nabla^2 \Omega) \,.
\end{align}    
\end{subequations}
%where we denote $\bar{G}_{2,X}$ as the effective mass squared $m^2$.
We vary this action with respect to the gauge-invariant quantities and obtain the tensor linear propagation equation,
\begin{align}
    \frac{1}{2}\bar{G}_4\Box h^{TT}_{ij}=0\,,
\end{align}
the linear vector equations,
\begin{align}
    (\bar{G}_{2,F}\Box-\bar{G}_{2,X})\Sigma_i=&0 \,,\\
    -\bar{G}_4\nabla^2 \Xi_i=&0\,,
\end{align}
and the linear scalar equations,
\begin{align}
    \bar{G}_4 (3\ddot{\Theta} - \nabla^2\Theta + 2\nabla^2 \Phi) =&0 \,,\\
    2\bar{G}_4\nabla^2 \Theta =&0 \,,\\
    \bar{G}_{2,F}\nabla^2 \ddot{\Upsilon}+\bar{G}_{2,F}\nabla^2 \dot{\Omega}+\bar{G}_{2,X}\nabla^2\Upsilon =& 0 \,,\\
    -\bar{G}_{2,F}\nabla^2\Omega-\bar{G}_{2,F}\nabla^2\dot{\Upsilon}+\bar{G}_{2,X}\Omega=&0 \,.
\end{align}
The equations above imply constraint equations on some of the variables, more specifically
\begin{align}\label{eqn:a_constr1_app}
    \Xi_i=0\,,\quad \Theta=0\,,\quad \Phi=0\,.
\end{align} 
As a consequence, the propagating vector equation can be written as 
\begin{align}
    (\Box - m^2)\Sigma_i=0 \,,
\end{align}
where we define the effective mass as 
\begin{align}
    m^2 := \bar{G}_{2,X}/\bar{G}_{2,F} \,.
\end{align}
Analogously, we substitute Eqs.~\eqref{eqn:a_constr1_app} into the linear scalar equations above and obtain a relation among the nonvanishing scalar modes,
\begin{align}
    \nabla^2 \Upsilon +\dot{\Omega}=0\,.
\end{align}
This is equivalent to imposing $\partial_\mu a^\mu = 0$, as already shown in the analytical derivation of Appendix~\ref{App:AnalyticSolCaseA}.
Finally, we can use this constraint to remove $\Omega$ and obtain the equation of motion for the dynamical mode $\Upsilon$,
\begin{align}
    (\Box - m^2)\Upsilon=0 \,,
\end{align}
where the effective mass is the same as the one defined for the vector mode.

Using the constraint equations above to remove the non-dynamical terms, the second-order action Eq.~\eqref{eq:a_sec_action} for the dynamical fields becomes
\begin{subequations}
    \begin{align}
    _{(2)}\mathcal{L}_T &= \frac{1}{4}\bar{G}_4 h^{TT}_{ij} \Box h_{TT}^{ij} \,,\\
    _{(2)}\mathcal{L}_V &= \frac{1}{2} \bar{G}_{2,F} \Sigma_{i} (\Box-m^2) \Sigma^{i}  \,,\\
    _{(2)}\mathcal{L}_S &= -\frac{1}{2}\bar{G}_{2,F}[ \nabla^2\Upsilon (\Box-m^2)\Upsilon ] \,.
\end{align}
\end{subequations}

\subsection{Case (b)}
Setting the effective mass to zero through the constraint $\bar{G}_{2,X}=0$, the second-order action Eq.~\eqref{eq:general_action} becomes
\begin{subequations}\label{eq:general_action_b}
\begin{align}
    {}_{\myst{(2)}}\mathcal{L}_\text{T}=-\frac{1}{4}(\bar{G}_4-\bar{A}^2\bar{G}_{4,X}){h}_{TT}^{ab}\ddot{h}^{TT}_{ab}+\frac{1}{4}\bar{G}_4 h_{TT}^{ab}\nabla^2 h^{TT}_{ab}\,,
\end{align}
\begin{align}
{}_{\myst{(2)}}\mathcal{L}_\text{V}
    &=\bar{G}_{2,F}(\frac{1}{2}\bar{A}^2 \Xi^i\Box\Xi_i+\bar{A}\Xi^i \Box\Sigma_i+\frac{1}{2}\Sigma^i\Box \Sigma_i) \nonumber\\
    & +\bar{G}_{2,Y}(-\bar{A}^4 \Xi^i \ddot \Xi_i-2\bar{A}^3 \Xi^i \ddot \Sigma_i - \bar{A}^2 \Sigma^i \ddot \Sigma_i) \nonumber \\
    &-\frac{1}{2}(\bar{G}_4+\bar{A}^2\bar{G}_{4,X})\Xi^i\nabla^2\Xi_i - \bar{A}\bar{G}_{4,X}\Xi^i\nabla^2\Sigma_i \,, 
\end{align}
\begin{align}
    {}_{\myst{(2)}}\mathcal{L}_\text{S}
    &=\bar{G}_{4,X} [\bar{A}^2(-\frac{3}{2}\Theta\ddot{\Theta}+2\Phi\nabla^2\Theta)-2\bar{A}\Theta\nabla^2(\Omega+\dot{\Upsilon})] \nonumber\\
    &+\bar{G}_4 (\frac{3}{2}\Theta \ddot{\Theta}-\frac{1}{2}\Theta\nabla^2\Theta +2\Phi\nabla^2\Theta) \nonumber\\
    &+\bar{G}_{3,X} [\frac{3}{2}\bar{A}^3\Theta\dot{\Phi}-\bar{A}^2(\frac{3}{2}\Theta\dot{\Omega}+\Upsilon\nabla^2\Phi)+\bar{A}\Upsilon\nabla^2 \Omega]\nonumber\\
    &+\bar{G}_{2,XX}(\frac{1}{2}\bar{A}^4\Phi^2 -\bar{A}^3\Phi\Omega +\frac{1}{2}\bar{A}^2\Omega^2) \nonumber\\
    & +(\bar{G}_{2,F}+2\bar{A}^2\bar{G}_{2,Y})(-\frac{1}{2} \dot{\Upsilon}\nabla^2 \dot{\Upsilon}-\frac{1}{2}\Omega\nabla^2 \Omega \nonumber\\
    &-\dot \Upsilon \nabla^2 \Omega -2\bar{A}^2 \Phi\nabla^2 \Phi + 2\bar{A}\Omega\nabla^2\Phi+2\bar{A}\dot{\Upsilon}\nabla^2\Phi) \,.
\end{align}
\end{subequations}

The linear equation of motion for the tensor field $h_{ij}^{TT}$ reads
\begin{align}
    -\frac{1}{2}(\bar{G}_4\Box + \bar{A}^2 \bar{G}_{4,X}\partial_0^2) h_{ij}^{\text{TT}} = 0\,.
\end{align}
When $\bar{G}_4 - \bar{A}^2\bar{G}_{4, X} = 0$, the tensor equation is not dynamical, so we only consider the physically interesting case when $\bar{G}_4 -\bar{A}^2\bar{G}_{4, X} \neq 0$. Given that $G_4 \neq 0$, a necessary assumption to have a dynamical metric theory of gravity, we can simplify the above expression into a plane wave equation for the tensor mode,
\begin{align}
-\ddot h_{ij}^{TT} +\frac{\bar{G}_4}{\bar{G}_4 - \bar{A}^2\bar{G}_{4, X} }\,\partial^2h_{ij}^{TT}  = 0\,.
\end{align}

The leading order equations for the vector DOFs $\Sigma$ and $\Xi$ are obtained from Eq.~\eqref{eq:general_action_b} as
\begin{subequations}
    \begin{align}
        0&=(\bar{G}_{2,F}+2\bar{A}^2\bar{G}_{2,Y}) \Box(\bar{A}\Xi_i +\Sigma_i) \nonumber\\
        &-2\bar{A}^2\bar{G}_{2,Y} \nabla^2 (\bar{A}\Xi_i +\Sigma_i) -\bar{A} \bar{G}_{4,X} \nabla^2 \Xi_i  \,,\\
        0&=\bar{A}(\bar{G}_{2,F}+2\bar{A}^2\bar{G}_{2,Y}) \Box (\bar{A} \Xi^j +\Sigma^j) \nonumber\\
        & - \bar{A} (\bar{G}_{4,X}+2\bar{A}^2 \bar{G}_{2,Y}) \nabla^2 (\bar{A} \Xi^j +\Sigma^j)  -\bar{G}_4\nabla^2 \Xi^j  \,. 
    \end{align}
\end{subequations}

\begin{comment}
\begin{subequations}
    \begin{align}
        0&=\bar{A}\Box(\bar{A}\Xi_i +\Sigma_i) -\bar{A}^2 \bar{G}_{4,X} \nabla^2 \Xi_i  \,,\\
        0&=\bar{A} \Box (\bar{A} \Xi^j +\Sigma^j)- \bar{A} \bar{G}_{4,X} \nabla^2 (\bar{A} \Xi^j +\Sigma^j) \nonumber \\
        & -\bar{G}_4\nabla^2 \Xi^j  \,. 
    \end{align}
\end{subequations}
\end{comment}

To find the propagation equations, we subtract $\bar{A}$ times the first equation from the second equation and obtain a constraint equation for $\Xi$, 
\begin{align}
    -\bar{A}\bar{G}_{4,X}\nabla^2 \Sigma_i - \bar{G}_4 \nabla^2 \Xi_i = 0,
\end{align}
\begin{comment}
\begin{align}
    \bar{G}_4 \nabla^2 \dot{\Xi}^i +\bar{A}\bar{G}_{4,X} \nabla^2 \dot{\Sigma}^i = 0 \,,
\end{align}
\end{comment}
thus
\begin{align}\label{eq:Xiconstrapp}
    \Xi_i = -\frac{\bar{G}_{4,X}}{\bar{G}_4}\bar{A} \Sigma_i= -\sigma\bar{A} \Sigma_i\,.
\end{align}

We substitute the vector constraint to simplify the dynamical equation for $\Sigma$, which turns out to be
\begin{align}
    0  =&  [(\bar{G}_{2,F}+2\bar{A}^2\bar{G}_{2,Y})(1-\sigma \bar{A}^2) \Box \nonumber\\
    & + (\sigma \bar{A}^2 \bar{G}_{4,X}-2\bar{A}^2 (1-\sigma \bar{A}^2)\bar{G}_{2,Y}) \nabla^2] \Sigma_i  \,.
\end{align}
\begin{comment}
\begin{align}
    [(1-\sigma \bar{A}^2) \Box  + \sigma \bar{A}^2 \bar{G}_{4,X} \nabla^2] \Sigma_i = 0 \,.
\end{align}
\end{comment}

If $1-\sigma \bar{A}^2=0$ 
%(the background $\bar{A}$ is a specific constant $\sqrt{1/\sigma}$) 
or $\bar{G}_{2,F}+2\bar{A}^2\bar{G}_{2,Y}=0$, the dynamical equation reduces to a Laplacian equation for $\Sigma$. Under the boundary condition that all fields vanish as $r \rightarrow +\infty$, this implies that $\Sigma_i = 0$, i.e. there is no vector mode.\newline
In the case where $(1-\sigma \bar{A}^2) \neq 0$ and $\gamma\neq 0$, we arrive at the following wave equation for the vector mode,
\begin{align}
    -\ddot \Sigma_i +\frac{\bar{G}_{2,F}(1-\sigma \bar{A}^2)+\sigma \bar{A}^2 \bar{G}_{4,X}}{\gamma(1-\sigma \bar{A}^2)}\,\partial^2\Sigma_i  = 0\,.
\end{align}

Finally, we vary the second-order action with respect to the gauge-invariant scalar variables $\Theta$, $\Phi$, $\Omega$ and $\Upsilon$, and we find respectively the following equations,

\begin{subequations}\label{eq:scalarEOMlin}
\begin{align}
    0= & \bar{G}_{4,X}[\bar{A}^2(-3\ddot\Theta + 2 \nabla^2 \Phi)- 2\bar{A} \nabla^2( \Omega + \dot \Upsilon)] \nonumber\\
    & + \bar{G}_4(3\ddot\Theta-\nabla^2 \Theta +2\nabla^2 \Phi) 
    + \frac{3}{2}\bar{G}_{3,X}(\bar{A}^3 \dot \Phi - \bar{A}^2 \dot \Omega)\,, \label{eq:linTheta}
\end{align}
\begin{align}
    0=& 2(\bar{A}^2 \bar{G}_{4,X} + \bar{G}_4) \nabla^2 \Theta + \bar{G}_{3,X}(-\frac{3}{2}\bar{A}^3 \dot \Theta -\bar{A}^2 \nabla^2 \Upsilon)\nonumber\\
    & +\bar{G}_{2,XX}(\bar{A}^4 \Phi -\bar{A}^3 \Omega)  \nonumber\\
    &+ 2\bar{A}(\bar{G}_{2,F}+2\bar{A}^2 \bar{G}_{2,Y})\nabla^2(-2\bar{A}\Phi + \Omega + \dot \Upsilon)\,, \label{eq:linPhi}
\end{align}
\begin{align}
    0=& -2\bar{A}G_{4,X}\nabla^2\Theta +\bar G_{3,X}\left(\frac{3}{2}\bar{A}^2\dot{\Theta}+\bar{A}\nabla^2\Upsilon\right) \nonumber\\
    & + \bar G_{2,XX}\left(-\bar{A}^3\Phi+\bar{A}^2\Omega \right)  \nonumber\\
    & +(\bar{G}_{2,F}+2\bar{A}^2 \bar{G}_{2,Y})\nabla^2\left(- \Omega -\dot{\Upsilon}  + 2\bar{A}\Phi \right)  \,, \label{eq:linOm} 
\end{align}
\begin{align}
    0= &2\bar{A} \bar G_{4,X}\nabla^2\dot{\Theta} +\bar{A} \bar G_{3,X}\nabla^2(-\bar{A}\Phi+\Omega) \nonumber \\
    & + (\bar{G}_{2,F}+2\bar{A}^2 \bar{G}_{2,Y})\nabla^2(\ddot{\Upsilon}+\dot{\Omega}-2\bar{A}\dot{\Phi}) \,. \label{eq:linPsi}
\end{align}
\end{subequations}
%%%%%%%
\begin{comment}
\begin{subequations}\label{eq:scalarEOMlin}
\begin{align}
    0&=\frac{3}{2}\bar{A}^2
\bar{G}_{3,X}(\bar{A}\dot{\Phi}-\bar{G}_{4}\nabla^2\Theta-\dot{\Omega})+2\bar{A}\bar{G}_{4,X}\nabla^2(\Omega-\dot{\Upsilon})\nonumber \\ 
&+3(\bar{G}_{4}-\bar{A}^2\bar{G}_{4,X})\ddot{\Theta}+2(\bar{G}_{4}-\bar{A}^2\bar{G}_{4,X})\nabla^2\Phi\,, \label{eq:linTheta} \\ 
    0&= 2\nabla^2(\dot{\Upsilon}-2\bar{A}\Phi+\Omega)+\bar{A}G_{3,X}(\frac{3}{2}\bar{A}\dot{\Theta}-\nabla^2\Upsilon)\nonumber\\
    &+2\left(\frac{G_{4}}{\bar{A}}+\bar{A}G_{4,X}\right)\nabla^2\Theta+\bar{A}^2G_{2,XX}(\bar{A}\Phi-\Omega)\,, \label{eq:linPhi}\\ 
    0&=\nabla^2(2\bar{A}\Phi-\dot{\Upsilon}-\Omega)+\bar{A}G_{3,X}\left(\frac{3}{2}\bar{A}\dot{\Theta}+\nabla^2\Upsilon\right)\nonumber \\   &-2\bar{A}G_{4,X}\nabla^2\Theta+\bar{A}^2G_{2,XX}\left(\frac{1}{4}\Omega-\bar{A}\Phi\right)\,,\label{eq:linOm} \\ 0&=\nabla^2(\ddot{\Upsilon}-2\bar{A}\dot{\Phi}+\dot{\Omega})+\bar{A}G_{3,X}\nabla^2(\Omega-\bar{A}\Phi)\nonumber \\
    &+2\bar{A}G_{4,X}\nabla^2\dot{\Theta}\,.\label{eq:linPsi}
\end{align}
\end{subequations}
\end{comment}
\newline
We sum Eqs.~\eqref{eq:linPhi} and $\bar A$ time Eq.~\eqref{eq:linOm} and express $\Phi$ in terms of the other scalar variables,
\begin{align}\label{eq:phiconstrapp}
    \Phi=\frac{\bar{G}_4}{\bar{A}^2\gamma}\Theta+\frac{1}{2\bar{A}}(\Omega+\dot\Upsilon)\,,
\end{align}
where the coefficient $\gamma$ is nonzero by the assumption that the vector mode is dynamical.
\newline
Next, we substitute this expression in Eq.~\eqref{eq:linOm} to obtain a constraint equation for $\Omega$,
\begin{comment}
    \begin{align}   \label{eq:omconstr}
\Omega&=\dot{\Upsilon} -\frac{2\bar{G}_{3,X}}{\bar{A}\bar{G}_{2,XX}}\nabla^2\Upsilon
+\frac{2\bar{G}_4}{\bar{A}\gamma}\Theta-\frac{3\bar{G}_{3,X}}{\bar{G}_{2,XX}}\dot{\Theta}\nonumber\\
 &-\frac{4(\bar{G}_4-\bar{A}^2\bar{G}_{4,X})}{\bar{A}^3\bar{G}_{2,XX}} \nabla^2 \Theta\,.
\end{align}
\end{comment}
\begin{align}   \label{eq:omconstrapp}
\bar{G}_{2,XX}\Omega&=\bar{G}_{2,XX}\dot{\Upsilon} -\frac{2\bar{G}_{3,X}}{\bar{A}}\nabla^2\Upsilon
+\bar{G}_{2,XX}\frac{2\bar{G}_4}{\bar{A}\gamma}\Theta\nonumber\\
 &-3\bar{G}_{3,X}\dot{\Theta}-\frac{4(\bar{G}_4-\bar{A}^2\bar{G}_{4,X})}{\bar{A}^3} \nabla^2 \Theta\,.
\end{align}
\newline
Multiplying Eq.~\eqref{eq:linPsi} with $\bar G_{2,XX}$, and plug in the above relations for $\Omega$ and $\Phi$ into it, we obtain a relation for $\Upsilon$,
\begin{comment}
   \begin{align}\label{eq:psiconstrs}
    \nabla^2\Upsilon&=-\frac{2(\bar{G}_4-\bar{A}^2\bar{G}_{4,X})}{\bar{A}^2\bar{G}_{3,X}}\nabla^2\Theta\nonumber \\
    &-\left(\frac{3\bar{A}}{2}+\frac{2(\bar{G}_4-\bar{A}^2\bar{G}_{4,X})\bar{G}_{2,XX}}{\bar{A}\bar{G}_{3,X}^2}\right)\dot{\Theta}\,.
\end{align} 
\end{comment}
\begin{align}\label{eq:psiconstrsapp}
    \bar{G}_{3,X}^2\nabla^2\Upsilon&=-\frac{2(\bar{G}_4-\bar{A}^2\bar{G}_{4,X})}{\bar{A}^2}\bar{G}_{3,X}\nabla^2\Theta\nonumber \\
    &-\left(\frac{3\bar{A}}{2}\bar{G}_{3,X}^2+\frac{2(\bar{G}_4-\bar{A}^2\bar{G}_{4,X})\bar{G}_{2,XX}}{\bar{A}}\right)\dot{\Theta}\,.
\end{align}
Finally, multiplying Eq.~\eqref{eq:linTheta} by $\bar G_{2,XX} \bar G_{3,X}^2$, then substituting Eqs.~\eqref{eq:phiconstrapp},~\eqref{eq:omconstrapp},~\eqref{eq:psiconstrsapp} into it, we find the following: 
\begin{align}
    \bar G_4 \bar G_{2,XX}
    \left[-(1-\bar{A}^2\sigma)(3\bar{G}_{3,X}^2 + 4\bar{G}_{2,XX}\bar{G}_4\frac{1-\sigma\bar{A}^2}{\bar{A}^2})\ddot\Theta \right. \nonumber \\
    \left. +\bar{G}_{3,X}^2 (-1+\frac{4\bar{G}_4}{\bar{A}^2\gamma})\partial^2\Theta\right]=0 \,,
\end{align}
We assumed $\bar G_4 \neq 0$ before. If in addition $\bar G_{2,XX} \neq 0$, the above equation can be organized as
%we substitute Eqs.~ \eqref{eq:phiconstr}, \eqref{eq:omconstr}, \eqref{eq:psiconstrs} into Eq. \eqref{eq:linTheta}, and we find the scalar wave equation
\begin{align}
    -\ddot\Theta +\frac{\bar{G}_{3,X}^2(-\bar{A}^2+4\bar{G}_4/\gamma)}{(1-\bar{A}^2\sigma)(3\bar{A}^2\bar{G}_{3,X}^2 + 4\bar{G}_{2,XX}\bar{G}_4(1-\sigma\bar{A}^2))}\partial^2\Theta =0\,.
    \label{theq}
\end{align}
One can check that setting $\bar G_{2,XX}=0$ in the action, we can obtain the following wave equation for the scalar mode,
\begin{align}
    \bar G_4 
    \left(-3\bar{G}_{3,X}^2(1-\bar{A}^2\sigma) \ddot\Theta  +\bar{G}_{3,X}^2 (-1+\frac{4\bar{G}_4}{\bar{A}^2\gamma})\partial^2\Theta\right)=0 \,,
\end{align}
which can also be simplified to Eq.~\eqref{theq} with $\bar G_{2,XX}=0$.
Therefore, Eq.~\eqref{theq} is a general equation for the scalar propagating degree of freedom, and it is non-dynamical if and only if $3\bar{A}^2\bar{G}_{3,X}^2 + 4\bar{G}_{2,XX}\bar{G}_4(1-\sigma\bar{A}^2)=0$. 
Otherwise, it follows the same asymptotic behaviour as previously described in the tensor and the vector sector.

%%%%%%%%%%%%%%%%%%%%%%%%%%%%%%%%%%%%%%%%%
\section{Displacement memory from the energy pulse perspective}\label{app:mem_en_pulse}

We present a useful and illuminating point-particle/energy-pulse approximation for the computations of gravitational displacement memory. This is a generalization of the discussion in Ref.~\cite{Heisenberg:2025tfh} to include the case of massive degrees of freedom.

To start, recall that the phase and group velocities for a plane wave with energy $\omega$ and wave vector $\vec k$ are generally defined as
\begin{align}\label{eq:def_velocities}
    V \equiv \frac{\omega}{k} \, , \quad \beta \equiv \frac{d\omega}{dk} \,,
\end{align}
with $k = |\vec k|$.
The associated four-momentum of the plane wave is 
\begin{align}
    p^\mu =(\omega, kn^i)= (\omega, \frac{\omega}{V}n^i)\,,
\end{align}
and its four velocity reads
\begin{align}
    \beta^\mu = (1, \beta n^i) \,.
\end{align}

The energy-momentum tensor of an ideal energy pulse is \cite{misner_gravitation_1973,Weinberg1972,sutherland2019energymomentumtensorfieldparticle}
\begin{align}
    t^{\mu\nu}_p (x') = p^\mu_p \beta^\nu_p \delta^{(3)}(\vec{x}'-\vec x_p) \,,
\end{align}
where $\vec x_p$ is the ``center of the energy pulse". Assuming it follows a radially outward trajectory, it is possible to write
\begin{align}
    x_p^i (t') = r_p(t')n^i \equiv \beta_p (t'-t_0)n^i \,,
\end{align}
where we define 
\begin{align}
    u_p = t_0 = t'-\frac{r_p}{\beta_p} \,.
\end{align}
Notice that $u_p$ is different from the retarded time of the energy pulse $u' = t'-r'/\beta_p$.
Then the energy density, energy current, momentum density, and momentum current of this pulse are
\begin{subequations}
\begin{align}
    t^{00}_p (x') &= \omega_p \delta^{(3)}(\vec{x}'-\vec x_p) \,, \\
    t^{0i}_p (x') &= \omega_p \beta^i \delta^{(3)}(\vec{x}'-\vec x_p) n^i\,, \\
    t^{i0}_p (x') &= \frac{\omega_p}{V_p} \delta^{(3)}(\vec{x}'-\vec x_p) n^i\,, \\
    t^{ij}_p (x') &= \omega_p \frac{\beta_p}{V_p} \delta^{(3)}(\vec{x}'-\vec x_p)n^i n^j\,,
\end{align}
\end{subequations}
with $n_i \equiv x'_i/r'$.
Switch to the spherical coordinate, and replace $r'$ and $r_p$ with the retarded time and $u_p$, the delta function reads
\begin{align}
    \delta^{(3)}(\vec x' -\vec x_p)=&\frac{1}{r'^2} \delta(r'-r_p(t')) \delta^{(2)}(\Omega'-\Omega_p) \nonumber\\
    =& \frac{1}{r'^2 \beta_p} \delta(u'-u_p) \delta^{(2)}(\Omega'-\Omega_p)\,.
\end{align}
Physically, the energy pulse should extend enough such that the energy-momentum tensor of its associated wave packet is definite. Thus, we smooth out the delta function by replacing $\delta(u'-u_p)$ with 
\begin{align}
    D(u'-u_p) := \frac{1}{\sqrt{2\pi}\Delta u'}e^{-(u'-u_p)^2/2\Delta u'^2} \,,
\end{align}
a Gaussian wave packet of width $\Delta u' \sim 1/f_L$.\footnote{Detail discussion in Ref.~\cite{Heisenberg:2025tfh}.}

The energy flux is connected to the energy density through
\begin{align}
    \frac{dE_p}{du'd\Omega'} = r'^2 \beta_p t_{00}^p = \omega_p D(u'-u_p)\delta^{(2)}(\Omega'-\Omega_p) \,,
\end{align}
and the momentum current reads
\begin{align}
    t_{ij}^p (x') = \frac{\omega_p }{r'^2 V_p} D(u'-u_p) \delta^{(2)}(\Omega'-\Omega_p) n_i n_j\,.
\end{align}
Therefore, the momentum flux as defined in Eq.~\eqref{eq:EmGenFinal} is now
\begin{align}\label{eq:momflux_p}
    F_p(u',\Omega') =\frac{\omega_p }{V_p} D(u'-u_p) \delta^{(2)}(\Omega'-\Omega_p) \,,
\end{align}
and it's related to the energy flux by 
\begin{align}
    F_p(u',\Omega')\equiv 
    \frac{1}{ V_p} \frac{dE_p}{du'd\Omega'} \,,
\end{align}
when the source is just an energy pulse.

%%%%%%%%%%%%%%%%%%%%%
\paragraph*{Case (a).}
Substituting this into Eq.~\eqref{eq:memory_a}, we see that the memory sourced by an energy pulse in the massive but Lorentz preserving case ($a$) is 
\begin{align}
    \delta h_{ij}^{TT}(x) =  \frac{\kappa_0}{2\pi r} &\int_{-\infty}^{u} du' \int_{S^2}d\Omega'  \, \frac{1}{V_p} \frac{dE_p}{du'd\Omega'}
   \beta_{p} \left[  \mathcal{V}_p n'_i n'_j\right]^\text{TT} \,.
\end{align}
Since in this case the propagating scalar and vector modes are dispersive, we have $V_p = 1/\beta_p$, the memory sourced by the additional Proca field is then
\begin{align}
    \delta h_{ij}^{TT}(x) =  & \frac{\kappa_0}{2\pi r} \int_{-\infty}^{u} du' \int_{S^2}d\Omega'  \,  \frac{dE_p}{du'd\Omega'} \beta_{p}^2
    \left[  \mathcal{V}_p n'_i n'_j\right]^\text{TT} \nonumber\\
    =& \frac{\kappa_0}{2\pi r}  \int_{S^2}d\Omega'  \,  \frac{dE_p (u)}{d\Omega'} \beta_p^2
    \left[  \frac{n'_i n'_j}{1- \beta_p \vec n' \cdot \vec n} \right]^\text{TT}\,,
\end{align}
which is the same as the memory formalism we derived for Horndeski \cite{Heisenberg:2024cjk}. 
When the energy flux is a delta function, the memory contribution of a monochromatic plane wave with frequency $\omega_p$ is 
\begin{align}
    \delta h_{ij}^{TT}(x) =  & \frac{\kappa_0}{2\pi r} \omega_p \beta_{p}^2 \tilde H(u-u_p) 
    \left[  \frac{n^p_i n^p_j}{1- \beta_p \vec n_p \cdot \vec n} \right]^\text{TT}  \,,
\end{align}
denoting $n^p_i \equiv n'_i(\Omega_p) $, and 
\begin{align}
    \tilde H(u-u_p) := \int_{-\infty}^u D(u'-u_p) du' \,,
\end{align}
a smoothed out step function with an effective high-frequency cutoff at $f_L$.

%%%%%%%%%%%%%%%%%%%%%
\paragraph*{Case (b).}
On the other hand, plugging in the momentum flux of an energy pulse [Eq.~\eqref{eq:momflux_p}] into the memory formula in the Lorentz-violating case (b) [Eq.~\eqref{eq:memory_b1},~\eqref{eq:memory_b2+},~\eqref{eq:memory_b2-}], we see that when $\beta_p \leq \beta_\text{T}$,
\begin{align}
    \delta h_{ij}^\text{TT} (x) = & \frac{\kappa_0}{2\pi r} \int_{-\infty}^{u} du' \int_{S^2}d\Omega'  \,  
    \beta_{p} \left[  \mathcal{V}_p(\Omega') n'_i n'_j\right]^\text{TT} \times\nonumber\\
    & \qquad \qquad \qquad \times \frac{\omega_p }{V_p} \delta(u'-u_p) \delta^{(2)}(\Omega'-\Omega_p) \nonumber\\
    =& \frac{\kappa_0}{2\pi r} \omega_p \tilde H(u-u_p) \left[  \mathcal{V}_p(\Omega_p) n^p_i n^p_j\right]^\text{TT}\,. 
\end{align}
The two velocities $\beta_p$ and $V_p$ cancel out because the propagating modes are non-dispersive in case (b).
If the velocity of the energy pulse exceeds the tensor mode velocity, $\beta_p >\beta_\text{T}$, depending on the direction of the pulse $\Omega_p$, the memory contribution is 
\begin{align}
    \delta h_{ij+}^\text{TT} (r,u,\Omega,\Omega_p) = \frac{\kappa_0 \omega_p}{2\pi r}  \tilde H(u-u_p) \left[  \mathcal{V}_p n^p_i n^p_j\right]^\text{TT}\,
\end{align}
when $\mathcal{V}_p(\Omega_p) = 1/(1-\frac{\beta_p}{\beta_\text{T}}\vec n_p \cdot \vec n) >0$,
and 
\begin{equation}
\delta h^{TT}_{ij-}(r,u,\Omega,\Omega_p)
=
\frac{\kappa_0\omega_p}{2\pi r}
\left[\widetilde H(u-u_p)-1\right]
\left[\mathcal V_p\,n_i^p n_j^p\right]^{TT}
\end{equation}
when $\mathcal V_p(\Omega_p)<0$. This is equivalent to the energy pulse memory formula derived in Ref.~\cite{Heisenberg:2025tfh}. In both cases, the total displacement-memory offset, defined as the difference between the late- and early-time values, therefore takes the common form
\begin{equation}
\Delta h^{TT}_{ij}
=
\frac{\kappa_0\omega_p}{2\pi r}
\left[
\mathcal V_p n_i^p n_j^p
\right]^{TT},
\qquad
\mathcal V_p\neq 0 .
\end{equation}

\begin{figure*}[t]
    \centering
    \subfigure{
        \includegraphics[width=0.42\textwidth]{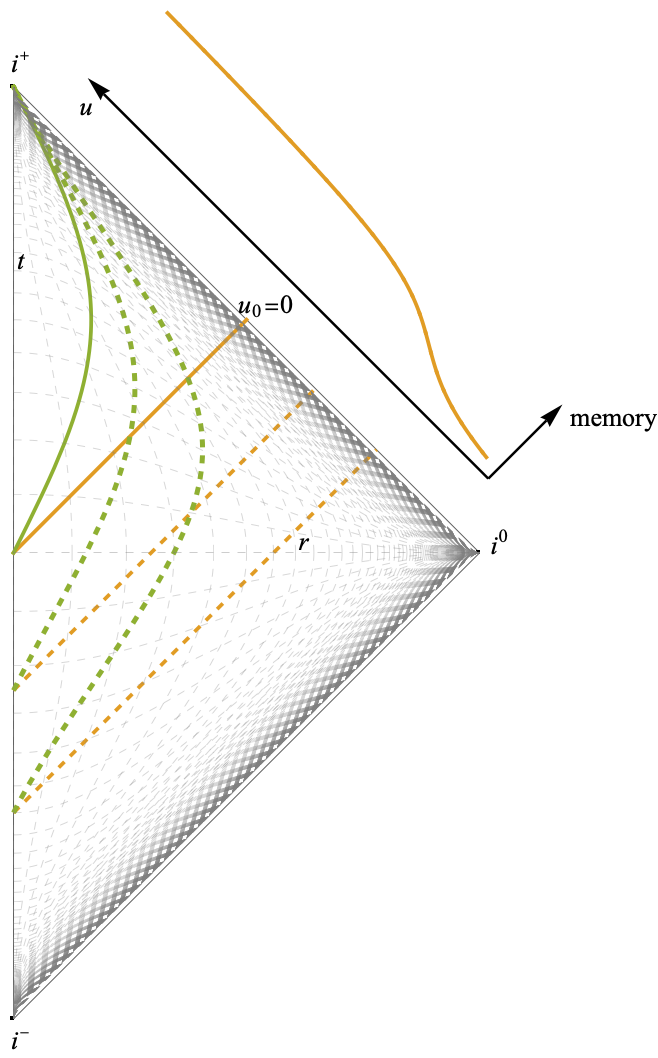}
        \label{fig:left}
    }
    \hfill
    \subfigure{
        \includegraphics[width=0.42\textwidth]{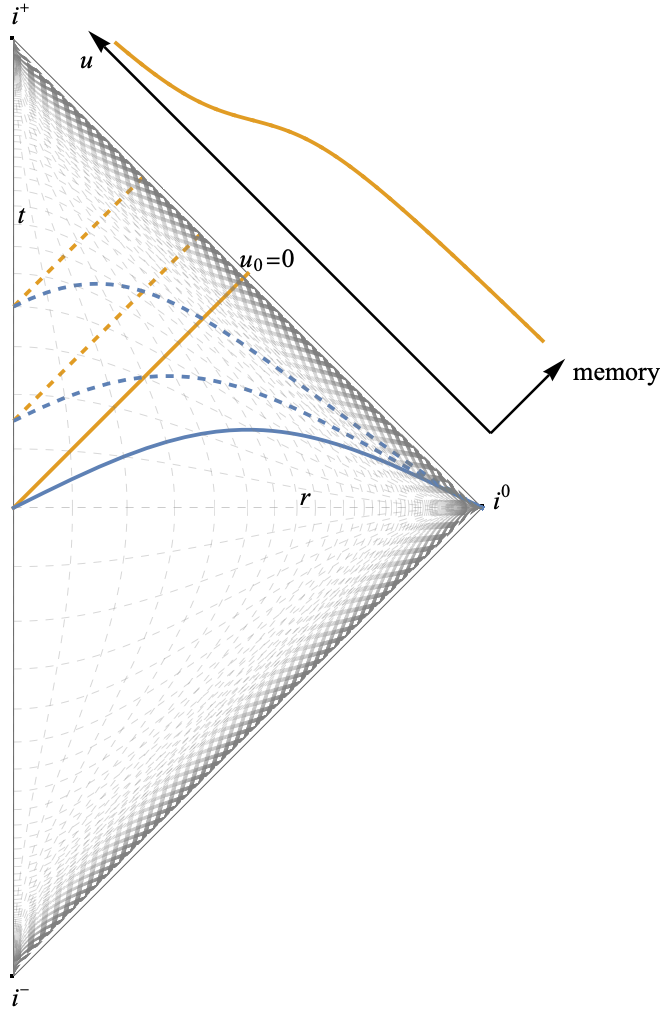}
        \label{fig:right}
    }
    \caption{The left panel shows the Penrose diagram for gravitational wave emission with a source propagating subluminally in the direction of the tensor modes propagation, while the right panel corresponds to a superluminally propagating source in this specific spatial direction. The curve on top of the diagrams shows how the memory accumulated in retarded time $u$ for each case.}
    \label{fig:penrosediagram}
\end{figure*}

Extending the discussion to multiple pulses, we can see how the memory detected at $u=u_0$ is affected by the other pulses. When the speed of the sources exceeds the speed of the tensor mode, the tensor memory measured at $u=u_0$ picks up the influence of all the sources emitted after $u_0$. In contrast, the tensor memory at $u_0$ was influenced by the earlier emissions if the source propagates subluminally, as shown in Fig.~\ref{fig:penrosediagram}.

\newpage
\bibliographystyle{utcaps}
\bibliography{references}

\clearpage

\end{document}